\documentstyle[12pt]{article}

\makeatletter
\@addtoreset{equation}{section}
\makeatother

\def\q{\,\qquad}
\def\NPB{{\em Nucl. Phys.} B}

\def\be{\begin{equation}}
\def\ee{\end{equation}}
\def\bea{\begin{eqnarray}}
\def\eea{\end{eqnarray}}
\newcommand{\hsa}{$cu(1,0|8)\,\,$}
\newcommand{\hsao}{$hu_0(1,0|8)\,\,$}

\newcommand{\N}{{\cal N}}
\newcommand{\D}{{\tilde{D}}}

\begin{document}


\begin{center}
{\large\bf Cubic Interactions of Bosonic Higher Spin Gauge Fields in $AdS_5$}
\vglue 0.6  true cm
\vskip1cm
{\bf M.A.~VASILIEV}
\vglue 0.3  true cm

I.E.Tamm Department of Theoretical Physics, Lebedev Physical Institute,\\
Leninsky prospect 53, 119991, Moscow, Russia
\vskip2cm
\end{center}

\begin{abstract}
The dynamics of totally symmetric free massless higher spin fields in
$AdS_d$ is reformulated in terms of the compensator formalism for AdS gravity.
The $AdS_5$  higher spin algebra is identified with the star product algebra
with the $su(2,2)$ vector (i.e., $o(4,2)$ spinor) generating elements.  Cubic
interactions of the totally symmetric bosonic higher spin gauge fields in
$AdS_5$, including the interaction with gravity, are formulated at the action
level.
\end{abstract}

\newcommand{\ty}{\hat{y}}
\newcommand{\bee}{\begin{eqnarray}}
\newcommand{\eee}{\end{eqnarray}}
\newcommand{\nn}{\nonumber}
\newcommand{\lis}{Fort1,FV1,LV}
\newcommand{\hy}{\hat{y}}
\newcommand{\by}{\bar{y}}
\newcommand{\bz}{\bar{z}}
\newcommand{\go}{\omega}
\newcommand{\e}{\epsilon}
\newcommand{\f}{\frac}
\newcommand{\p}{\partial}
\newcommand{\half}{\frac{1}{2}}
\newcommand{\ga}{\alpha}
\newcommand{\gal}{\alpha}
\newcommand{\U}{\Upsilon}
\newcommand{\C}{{\bf C}}
\newcommand{\ups}{\upsilon}
\newcommand{\bu}{\bar{\upsilon}}
\newcommand{\dga}{{\dot{\alpha}}}
\newcommand{\dgb}{{\dot{\beta}}}
\newcommand{\gb}{\beta}
\newcommand{\gga}{\gamma}
\newcommand{\gd}{\delta}
\newcommand{\gl}{\lambda}
\newcommand{\gk}{\kappa}
\newcommand{\gep}{\epsilon}
\newcommand{\gvep}{\varepsilon}
\newcommand{\gs}{\sigma}
\newcommand{\V}{|0\rangle}
\newcommand{\ws}{\wedge\star\,}
\newcommand{\gee}{\epsilon}
\newcommand{\ggg}{\gamma}
\newcommand\ul{\underline}
\newcommand\un{{\underline{n}}}
\newcommand\ull{{\underline{l}}}
\newcommand\um{{\underline{m}}}
\newcommand\ur{{\underline{r}}}
\newcommand\us{{\underline{s}}}
\newcommand\up{{\underline{p}}}

\newcommand\uq{{\underline{q}}}
\newcommand\ri{{\cal R}}
\newcommand\punc{\multiput(134,25)(15,0){5}{\line(1,0){3}}}
\newcommand\runc{\multiput(149,40)(15,0){4}{\line(1,0){3}}}
\newcommand\tunc{\multiput(164,55)(15,0){3}{\line(1,0){3}}}
\newcommand\yunc{\multiput(179,70)(15,0){2}{\line(1,0){3}}}
\newcommand\uunc{\multiput(194,85)(15,0){1}{\line(1,0){3}}}
\newcommand\aunc{\multiput(-75,15)(0,15){1}{\line(0,1){3}}}
\newcommand\sunc{\multiput(-60,15)(0,15){2}{\line(0,1){3}}}
\newcommand\dunc{\multiput(-45,15)(0,15){3}{\line(0,1){3}}}
\newcommand\func{\multiput(-30,15)(0,15){4}{\line(0,1){3}}}
\newcommand\gunc{\multiput(-15,15)(0,15){5}{\line(0,1){3}}}
\newcommand\hunc{\multiput(0,15)(0,15){6}{\line(0,1){3}}}
\newcommand\ls{\!\!\!\!\!\!\!}

\section{Introduction}

Irreducible relativistic
fields in the flat $d$-dimensional space-time classify according to
the finite-dimensional representations of the
Wigner little algebra $l$. It is well known that $l=o(d-2)$ for
the massless case $m=0$ and $l=o(d-1)$ for  $m\neq 0$.
{}From the field-theoretical viewpoint the difference between the massless
and massive cases is that, except for the scalar and
spinor fields, all massless fields possess specific gauge symmetries
reducing a number of independent degrees of freedom.

Since the totally  antisymmetric symbol $\epsilon^{a_1 \ldots a_n}$
$(a=1\div n)$ is  $o(n)$ invariant it is enough to
consider the representations of $o(n)$ associated with the
Young diagrams having at most $\left [\half n\right ]$ rows.
For lower dimensions like $d=4$ and $d=5$ only the
totally symmetric massless higher spin representations of the little algebra
appear, characterized by a single number $s$. An
integer spin $s$ massless field is described by a totally
symmetric tensor $\varphi_{n _1\ldots n_s}$ subject to the double
tracelessness condition \cite{Fr} $\varphi ^{r}{}_{r}{}^{s}{}_{s n
_5\ldots n_s}=0$ which is nontrivial for $s\geq 4$. A quadratic action
\cite{Fr} for a free spin $s$ field $\varphi _{n _1\ldots n_s}$
is fixed unambiguously \cite{C} up to an overall factor in the form
\begin{eqnarray}
\label{fract}
S_s&=&\frac{1 }{2 }(-1)^s
\int d^4 x\,\{\partial_n\varphi_{m_1\ldots m_s}\partial
^n\varphi ^{m_1 \ldots m_s}\\
&{}&\ls\ls\ls-\frac{1 }{2}s(s-1)\partial_n\varphi^r {}_{r m_1 \ldots
m_{s-2}}\partial^n\varphi^k{}_k{}^{m_1\ldots
m_{s-2}}\nonumber
+s(s-1)\partial_n\varphi^r{}_{r m_1\ldots m_{s-2}}\partial_k\varphi
^{n k m_1\ldots m_{s-2}}\\
&{}&\ls\ls\ls - s\partial_n\varphi^n{}_{m_1\ldots m_{s-1}}\partial
_r\varphi^{r m_1\ldots m_{s-1}} - \frac{1 }{4
}s(s-1)(s-2)\partial_n\varphi^r{}_r{}^n{}_{m_1\ldots m
_{s-3}}\partial_k\varphi^t{}_t{}^{k m_1\ldots m_{s-3}}\}\,
\nonumber
\end{eqnarray}
by the requirement of gauge invariance
under the Abelian gauge transformations
\begin{equation}
\delta\varphi_{n_1\ldots n_s}=\partial_{\{n_1}\varepsilon_{n_2\ldots n_{s}\}}
\end{equation}
with the parameters $\varepsilon_{n _1...n _{s-1}}$ being rank $(s-1)$
totally symmetric traceless tensors,
$\varepsilon^{r}{}_{r n_3\ldots n_{s-1}}=0$.
This formulation is parallel \cite{WF} to the
metric formulation of gravity and is called
formalism of symmetric tensors.
Fermionic higher spin gauge fields are described analogously \cite{FFr}
in terms of rank-$(s-1/2)$ totally
symmetric spinor-tensors $\psi _{n _1...n _{s-1/2}\alpha }$
subject to the $\gamma -$tracelessness condition $\gamma
^s {}_{\alpha }  {}^{\beta } \psi ^r{} _{r s n _4...n
_{s-1/2}\beta }=0$. A progress on the covariant
description of generic (i.e., mixed symmetry)
massless fields in any dimension was achieved in \cite{lab,BPT}.

Higher spin gauge symmetry principle is the fundamental concept
of the theory of higher  spin massless fields.
By construction, the class of higher spin gauge theories
consists of most symmetric theories
having as many as possible symmetries unbroken\footnote{We
only consider the case of relativistic fields that upon quantization
are described by lowest weight unitary representations (lowest weight
implies in particular, that the energy operator is bounded from below).
Beyond this class some other ``partially massless" higher spin gauge
fields can be introduced \cite{DW} which are either non-unitary or
live in the de Sitter space (recall that de Sitter group $SO(d,1)$
does not allow lowest weight unitary representations).}.
Any more symmetric theory will
have more lower and/or higher spin symmetries and therefore
will belong to the class of higher spin theories.
As such, higher spin gauge theory is of particular importance for
the search of a fundamental symmetric phase of the superstring theory. This is
most obvious in the context of the so-called Stueckelberg
symmetries in the string field theory which have a form of some spontaneously
broken higher spin gauge symmetries. Whatever a symmetric phase of the
superstring
theory is, Stueckelberg symmetries are expected to become unbroken higher
spin symmetries in such a phase and, therefore, the superstring field
theory has to identify with one or another version of the
higher spin gauge theory.

The problem is to introduce interactions of  higher spin fields
with some other fields in a way compatible with  the
higher spin gauge symmetries. Positive results in this direction  were
first obtained for interactions
of higher spin gauge fields in the flat space
with the matter fields and with themselves but not with
gravity \cite{pos}. In the framework of gravity, the nontrivial higher
spin gauge theories were so far elaborated \cite{FV1,FV2,more}
(see also \cite{R1,gol} for reviews)
for $d=4$ which is the simplest nontrivial case
since higher spin gauge fields do not propagate if $d<4$.
As a result, it was found out that

\noindent
(i) in the framework of gravity, unbroken
higher spin gauge symmetries require a non-zero cosmological constant;

\noindent
(ii) consistent higher spin theories contain infinite sets of
infinitely increasing spins;

\noindent
(iii) consistent higher spin gauge
interactions contain higher derivatives: the higher spin is the more
derivatives appear;

\noindent
(iv) the higher spin symmetry algebras
\cite{FVa} identify
with certain star product algebras with spinor generating elements
\cite{Fort2}.

Some of these properties, like the relevance of the $AdS$
background and star product algebras, discovered in the eighties
were rather unusual at that time but got their
analogues in the latest superstring developments
in the context of AdS/CFT correspondence \cite{AdS/CFT} and
the non-commutative Yang-Mills limit \cite{ncYM}.
We believe that this convergency  can unlikely be occasional.
Let us note that recently an attempt to incorporate the
dynamics of higher spin massless into the two-time version of
the non-commutative phase space approach was undertaken in
\cite{BD}.

The feature that unbroken
higher spin gauge symmetries require a non-zero cosmological constant
is of crucial importance in several respects. It
explained why negative conclusions on the existence of the consistent
higher-spin-gravitational interactions were obtained in \cite{diff}
where the problem was analyzed within an expansion near the flat
background. Also it explains why the higher spin gauge theory phase is not
directly seen in the M theory (or superstring theory) framework prior
its full formulation in the $AdS$ background is achieved.
The same property makes the $S$-matrix
Coleman-Mandula-type no-go arguments \cite{cm}
irrelevant because there is no $S$-matrix in the $AdS$ space.

A challenging problem of the higher spin gauge theory is to extend
the $4d$ results on the higher-spin-gravitational interactions
to higher dimensions. This is of particular
importance  in the context of the possible applications of the
higher spin gauge theory to the superstring theory ($d=10$) and
M theory (d=11).
A conjecture on
the structure of the higher spin symmetry algebras
in any dimension  was made in \cite{vf}
where the idea was put forward that analogously to what was proved to be true
in $d=4$ \cite{Fort2} and $d<4$ (see \cite{gol} for references)
higher spin algebras in any dimension are certain star product algebras
with spinor generating elements.

As a first step towards higher dimensions it is illuminating to
analyze the next to $d=4$ nontrivial case, which
is $d=5$. This is the primary goal of this paper.
The case of $AdS_5/CFT_4$ higher spin duality  is particularly
interesting in the context of duality of the type IIB superstring
theory on $AdS_5\times S^5$ with a constant Ramond-Ramond field strength
to the $\N=4$ supersymmetric Yang-Mills theory \cite{AdS/CFT}.
It was conjectured recently in \cite{W,Sun}
that the boundary theories dual to the
$AdS_{d+1}$ higher spin gauge theories are free conformal theories
in $d$ dimensions.  This conjecture is in agreement with the results of
\cite{KVZ} where the conserved conformal higher spin currents
bilinear in the
$d$-dimensional free massless scalar field theory were shown to be in
the one-to-one correspondence with the set of the $d+1$-dimensional
bulk higher spin gauge fields
associated with the totally symmetric representations of the little
algebra. In contrast to the regime $g^2 n\to \infty $ underlying
the standard AdS/CFT correspondence \cite{AdS/CFT}, as conjectured in
\cite{W,Sun}, the AdS/CFT regime associated with the higher spin gauge
theories corresponds to the limit $g^2 n\to 0$. The realization of the
higher spin conformal symmetries in the
$4d$ free boundary conformal theories was considered recently
in \cite{BHS}. It was shown that they indeed possess the
global higher spin conformal symmetries proposed long ago
by Fradkin and Linetsky \cite{FLA} in the context of
$4d$ conformal higher spin gauge theories \cite{FL}, and some their
further extensions\footnote{The conformal
higher spin gauge theories of \cite{FL} generalize $C^2$  Weyl
gravity and  are non-unitary because of the higher derivatives
in the kinetic terms that give rise to ghosts.}.
It was conjectured in \cite{BHS} that the $4d$ conformal higher
spin gauge symmetries can be realized as higher spin
gauge symmetries of $AdS_5$ bulk unitary higher spin gauge theories.
Analogous conjecture was made in \cite{SSd} with respect to the minimal
infinite-dimensional reduction of the $4d$ conformal higher spin algebra.

The $AdS_5$ case is more complicated compared to
 $AdS_4$. Naively, one might think that only
one-row massless higher spin representations
of the little group characterized by a single number $s$ appear.
However, there is a catch due to the fact that the classification
of massless fields in $AdS_d$ is different \cite{BMV} from that of
the flat space. As a result, more types of massless fields appear
in $AdS_5$ which all reduce in the flat limit to the some combinations
of the symmetric fields. In the $AdS_5$ case however, they are expected to
be described by the dynamical fields having the symmetry
properties of the two row Young diagrams. Unfortunately, so far the
covariant formalism for the description of such fields in
the $AdS$ space, that would extend that developed in
\cite{LV,VF} for the totally symmetric higher spin fields, was
not worked out. This complicates a formulation of the $AdS_5$
higher spin gauge theories for the general case.
In particular, based on the results presented in section
\ref{$su(2,2)$ - $o(4,2)$ Dictionary} of this paper, it was argued
in \cite{BHS} that such mixed symmetry higher gauge spin fields
have to appear in the $5d$ higher spin algebras with $\N \geq 2$
supersymmetries. For this reason, in this paper
we confine ourselves to the simplest purely bosonic $\N=0$ case
of the $AdS_5$ higher spin gauge theory. The $\N=1$ case will be
considered in the forthcoming paper \cite{AV}. To proceed beyond
$\N=1$ one has first of all to develop the appropriate formulation
of the $AdS_5$ massless higher spin fields that have the symmetry
properties of the two-row Young diagrams.

The organization of the paper is as follows. To make it as much
selfcontained as possible we start in section
\ref{Generalities} with a summary of the general features
of the approach developed in \cite{\lis} relevant to
the analysis of the $5d$ higher spin gauge theory in this paper.
In particular,
the main idea of the higher spin extension at the algebraic and Lagrangian
level is discussed in section \ref{Higher Spin Extension}.
In section \ref{Totally Symmetric Bosonic Massless Fields in
any Dimension} we summarize the main results of \cite{LV}
on the formulation of the totally symmetric bosonic massless fields
in any dimension, introducing covariant notation
based on the compensator approach to $AdS_d$ gravity explained in the
section \ref{$AdS_d$ Gravity with Compensator}.
In section \ref{Spinor Notation and $AdS_5$ Gravitational Field}
we reformulate $AdS_5$ gravity in the $su(2,2)$ spinor notations.
The correspondence between finite-dimensional representations of
$su(2,2)$ and $o(4,2)$ relevant to the $AdS_5$ higher spin problem
is presented in the
section \ref{$su(2,2)$ - $o(4,2)$ Dictionary}.
$AdS_5$ higher spin gauge algebras are defined in section
\ref{5d Higher Spin Algebra}. $su(2,2)$ systematics of the $5d$
higher spin massless is given in
 section \ref{d5 Higher Spin Gauge Fields}.
The unfolded form of the free equations of motion for all massless
totally symmetric tensor fields in $AdS_5$ called
Central On-Mass-Shell Theorem is presented in the
section \ref{5d Central On-Mass-Shell Theorem}.
The analysis of the $AdS_5$ higher spin action is the content of
section \ref{Higher Spin Action} where we, first, discuss some
general properties of the higher spin action, and then derive
the quadratic (section \ref{Quadratic Action})
and cubic (section \ref{Cubic Interactions})
higher spin actions possessing necessary higher spin symmetries.
The reductions to the higher spin gauge theories that describe
finite collections of massless fields of any given spin are
defined in section \ref{Maximally Reduced Model}.
Conclusions and some open problems are discussed
in section \ref{Conclusion}.
The Appendix contains a detailed derivation of the
the $5d$ free higher spin equations of motion.

\newpage

\section{Generalities}
\label{Generalities}

\subsection{$AdS_d$ Gravity with Compensator}
\label{$AdS_d$ Gravity with Compensator}

It is well-known that gravity admits a formulation in terms
of the gauge fields associated with one or another space-time
symmetry algebra \cite{ggr,MM,SW}.
Gravity with the cosmological term
in any space-time dimension  can be described
in terms of the gauge fields
$w^{AB}=-w^{BA}= dx^\un w_\un^{AB}$ associated with the $AdS_d$ algebra
$h=o(d-1,2)$ with the basis elements
$t_{AB}$. Here the underlined indices
$\um,\un\ldots = 0\div d-1$ are (co)tangent for the space-time base manifold
while $A,B = 0\div d$ are (fiber)
vector indices of the gauge algebra $h=o(d-1,2)$. Let $r^{AB}$ be the
Yang-Mills $o(d-1,2)$ field strength
\be
\label{r}
r^{AB} = dw^{AB}
+w^{AC}\wedge w_C{}^{B} \,,
\ee
where $d=dx^\un
\frac{\partial}{\partial x^\un}$ is the exterior differential.
One can use the decomposition
\be
w = w^{AB} t_{AB} = \go^{L\,ab} L_{ab} + \lambda e^a P_a\,
\ee
($a,b = 0\div d-1$). Here $\go^{L\,ab}$
is the Lorentz connection associated with the Lorentz subalgebra $o(d-1,1)$.
The frame field $e^a$ is associated
with  the $AdS_d$ translations $P_a$ parametrizing $o(d-1,2)/o(d-1,1)$.
Provided that $e^a$ is nondegenerate, the zero-curvature condition
\be
\label{rvac}
r^{AB} (w ) =0
\ee
implies that  $\go^{L\, ab}$ and $e^a$ identify with the gravitational
fields of $AdS_d$. $\lambda^{-1}$ is the radius of the $AdS_d$
space-time. (Note that $\lambda$
has to be introduced to make the frame field $e^a$ dimensionless.)

One can make these definitions covariant with the help of the
compensator field \cite{SW} $V^A (x)$ being a time-like $o(d-1,2)$
vector $V^A$ normalized to
\be
\label{vnorm}
V^A V_A=1
\ee
(within the mostly minus signature).
The Lorentz algebra then identifies with the stability subalgebra
of $V^A$. This allows for the covariant definition of the frame field and
Lorentz connection \cite{SW,PV}
\be
\label{defh}
\lambda E^A = D(V^A) \equiv d V^A + w^{AB}V_B\,,
\ee
\be
\go^{L\,AB} = w^{AB} - \lambda ( E^A V^B - E^B V^A )\,.
\ee
According to these definitions
\be
\label{ort}
E^A V_A =0 \,,
\ee
\be
D^L V^A = dV^A + \go^{L\,AB}V_B \equiv 0\,.
\ee
$V_A$ is the null vector of
$E^A=dx^\un E_\un^A$.
When the matrix $E_\un^A$ has the maximal rank $d$ it
can be identified with the frame field giving rise to
the nondegenerate space-time metric tensor
\be
g_{\un\um} = E_\un^A E_\um^B \eta_{AB}\,.
\ee

The torsion 2-form is
\be
t^A\equiv DE^A\equiv \lambda^{-1} r^{AB} V_B \,.
\ee
(Note that due to (\ref{ort}) $DE^A = D^L E^A$.)
The zero-torsion condition
\be
\label{0tor}
t^A = 0\,
\ee
expresses the Lorentz connection via
(derivatives of) the frame field in a usual manner.

With the help of $V_A$ it is straightforward to build a $d$--dimensional
generalization
of the $4d$ MacDowell-Mansouri-Stelle-West
pure gravity action
\cite{MM,SW}
\be
\label{gact}
S=-\frac{1}{4\lambda^2 \kappa^{d-2}}\int_{M^d}
\gep_{A_1 \ldots A_{d+1}}r^{A_1 A_2}\wedge r^{A_3 A_4}
\wedge E^{A_5}\wedge\ldots     \wedge E^{A_{d}} V^{A_{d+1}}\,.
\ee
Taking into account that
\be
\delta r^{AB} = D\delta w^{AB}\,,\qquad
\delta E^A  = \lambda^{-1}(\delta w^{AB} V_B +D \delta V^A )\,,\qquad
V_A \delta V^A =0 \,
\ee
along with the identity
\be
\label{eid}
\epsilon^{A_1 \ldots A_{d+1}} = V^{A_1} V_B
\epsilon^{B A_2 \ldots A_{d+1}}  +\ldots+  V^{A_{d+1}}V_B
\epsilon^{ A_1 \ldots A_{d} B}
\ee
one finds
\bee
\label{vgact}
\delta S&=&-\frac{1}{4\lambda^2 \kappa^{d-2}}\int_{M^d}
\gep_{A_1 \ldots A_{d+1}} r^{A_1 A_2}\wedge
\Big (2(-1)^d\lambda \delta w^{A_3 A_4}
\wedge E^{A_5} \wedge\ldots  \wedge E^{A_{d+1}}\nn\\
&+&(d-4)\lambda^{-1} r^{A_3 A_4}
\wedge\delta w^{A_5 B}V_B\wedge E^{A_6}\wedge \ldots
\wedge E^{A_{d}} V^{A_{d+1}} \Big )
+\delta_1 S       \,,
\eee
where
\bee
\label{vgpract}
\delta_1 S\!\!&=&\!\!-\frac{1}{4\lambda^2 \kappa^{d-2}}\int_{M^d}
\gep_{A_1 \ldots A_{d+1}} r^{A_1 A_2}\wedge t^{A_3}\wedge
\Big ((d-4)(2 \delta w^{A_4 A_5}
 \wedge E^{A_6} \nn\\
&+& (d-5)\lambda^{-1} r^{A_4 A_5}
\wedge \delta V^{A_6}) \wedge  E^{A_7}\wedge \ldots
\wedge E^{A_{d}}  V^{A_{d+1}}\nn\\
&+& 4(d-3) \lambda V^{A_4} \wedge E^{A_5}\wedge \ldots
\wedge E^{A_{d}} \delta V^{A_{d+1}} \Big )\,
\eee
is the part of the variation that contains torsion.

We shall treat the action $S$
perturbatively with $r^{AB}$ being small. According to
(\ref{rvac}) this implies a perturbation expansion around the $AdS_d$
background. In this framework, the second term
in (\ref{vgact})  and the first term in (\ref{vgpract}) only
 contribute to the nonlinear corrections
of the field equations for the gravitational fields $w^{AB}$.

For the part of  $\delta w^{AB}$ orthogonal to $V^C$
\be
\label{ortdo}
\delta w^{AB}=\delta \xi^{AB}\,,\qquad \delta \xi^{AB}V_B =0 \,
\ee
we obtain
\be
\label{loreq}
\frac{\delta S}{\delta \xi^{A_1 B_2}} =
{\kappa^{2-d}}
 \gep_{A_1 \ldots A_{d+1}} t^{A_3}\wedge \Big
( E^{A_4} \wedge E^{A_5}
-\frac{d-4}{2\lambda^2}r^{A_4 A_5}\Big )
 \wedge E^{A_6}\wedge \ldots \wedge E^{A_{d}} V^{A_{d+1}} \,.
\ee
Perturbatively, (i.e. for $r^{AB}$ small) (\ref{loreq})
is equivalent to the zero-torsion condition (\ref{0tor}).
In what follows we
will use the so called 1.5 order formalism. Namely, we will
assume that the zero-torsion constraint is imposed to express
the Lorentz connection via derivatives of the frame field.
The same time we will
use an opportunity to choose any convenient form for the
variation of the Lorentz connection because any  term
containing this variation is zero by (\ref{loreq}) and (\ref{0tor}).

The generalized Einstein equations originating from the variation
\be
\delta w^{AB} =\lambda ( \delta \xi^A V^B - \delta\xi^B V^A )
\,,\qquad \delta \xi^A V_A =0\,
\ee
are
\be
\label{geeq}
 \kappa^{2-d}
\gep_{A_1 \ldots A_{d+1}} r^{A_2 A_3}\wedge
\Big (
E^{A_4} \wedge E^{A_5}
-\frac{d-4}{4\lambda^{2}} r^{A_4 A_5}\Big )
\wedge E^{A_6}\wedge \ldots
\wedge E^{A_{d}} V^{A_{d+1}}=0\,.
\ee
The first term is nothing but the left-hand-side of the
Einstein equations with the cosmological  term. The second term
describes some additional interaction terms bilinear in the
curvature $r^{AB}$. These terms do not contribute to the
linearized field equations.
In the $4d$ case the additional terms are absent because the
corresponding part of the action is topological having the
Gauss-Bonnet form. Note that the additional
interaction terms contain higher derivatives together with the
 factor of $\lambda^{-2}$ that
diverges in the flat limit $\lambda\to 0$. Terms of this type
play an important role in the higher spin theories to guarantee
the higher spin gauge symmetries.
Let us note that the form of the equation (\ref{geeq}) indicates
that beyond $d=4$ the action (\ref{gact}) may have
other symmetric vacua\footnote{I
am grateful to K.Alkalaev for the useful discussion of this point.}
(e.g. with
$r^{AB} = \frac{4\lambda^2}{d-4} E^A\wedge E^B$).
We shall not discuss this point here in more detail because
its analysis requires the non-perturbative knowledge of the
higher spin theory which is still lucking for $d>4$.

{}From  (\ref{vgpract}) it follows that the
variation of $S$ with respect to the compensator $V^A$ is
proportional to the torsion 2-form $t^A$.
This means that, at least
perturbatively, there exists such a variation of the fields
\be
\label{adtr}
\delta V^A = \epsilon^A (x)\,, \qquad
\delta w^{AB} = \eta^{AB} (r,\epsilon )
\ee
with $\epsilon^A V_A =0$ and
some $\eta^{AB} (r,\epsilon )$ bilinear in $r^{AB}$ and $\epsilon^A$
that $S$ remains invariant. As a result,
there is an additional gauge symmetry that allows to
gauge fix $V^A$ to any value satisfying (\ref{vnorm}).
It is therefore shown  that $V^A$ does not carry
extra degrees of freedom.

The compensator field $V^A$ makes the $o(d-1,2)$ gauge  symmetry
manifest
\be
\delta w^{AB} = D \epsilon^{AB}  \,,\qquad
\delta V^A =- \epsilon^{AB}V_B\,.
\ee
Fixing a particular value of $V^A$ one is left with the
mixture of the gauge transformations that leave $V^A$ invariant, i.e.
with the parameters satisfying
\be
\label{leftsym}
0= \delta V^A = \epsilon^A (x) - \epsilon^{AB}V_B\,.
\ee
Since the additional transformation (\ref{adtr}) contains dependence
on the curvature $r^{AB}$, this property is inherited by the
leftover symmetry with the parameters satisfying (\ref{leftsym}).

The fact that there is an additional
symmetry (\ref{adtr}) is not a big surprise in the framework of
the theory of gravity formulated in terms of differential forms,
having explicit invariance under diffeomorphisms.
That this should happen is most clear from the
observation that the infinitesimal
space-time diffeomorphisms induced by  an arbitrary vector field
$\epsilon^\un (x)$ admit a representation
\be
\label{difg}
\delta w_\um^{AB} =\epsilon^\un  \partial_\un w^{AB}_\um + \partial_\um
(\epsilon^\un ) w^{AB}_\un =
\epsilon^\un r_{\un\um}{}^{AB} - D_\um \epsilon^{AB}\,,
\ee
\be
\delta V^A = \epsilon^\un  \partial_\un  V^A =
\epsilon^\un E_\un^A + \epsilon^{AB} V_B \,,
\ee
where
\be
\epsilon^{AB} = -\epsilon^\un w_\un{}^{AB}\,.
\ee
The additional gauge transformation (\ref{adtr})
with $\epsilon^A = \epsilon^\un E_\un^A$ can therefore be
understood as a mixture of the diffeomorphisms
and $o(d-1,2)$ gauge transformations.

Another useful interpretation of the formula (\ref{difg})
is that, for the vacuum solution satisfying (\ref{rvac}), diffeomorphisms
coincide with some gauge transformations. This observation
explains why the space-time symmetry algebras associated with
the motions of the most symmetric vacuum spaces reappear as gauge
symmetry algebras in the ``geometric approach" to gravity and its
extensions.

\subsection{General Idea of the Higher Spin Extension}
\label{Higher Spin Extension}

The approach to the theory of interacting higher spin gauge
fields developed originally
in \cite{Fort1,FV1} for the $d=4$ case
is a generalization of the ``geometric" approach to gravity sketched
in section \ref{$AdS_d$ Gravity with Compensator}.
The idea is to describe the higher spin gauge fields in terms of
the Yang-Mills gauge fields and field strengths
associated with an appropriate higher spin symmetry
algebra $g$  being some infinite-dimensional
extension of the finite-dimensional $AdS_d$ space-time symmetry
algebra $h=o(d-1,2)$.

Let the 1-form $\go (x)= dx^\un \go_\un (x)$ be
the gauge field of $g$ with the field strength (curvature 2-form)
\be
R= d\go + \go  \wedge * \go \,,
\ee
where $*$ is some associative product law
leading to the realization of $g$ via commutators. (This is analogous
to the matrix realization of the classical Lie algebras with the star
product instead of the matrix multiplication. A particular realization
of the star product relevant to the $5d$ higher spin dynamics is
given in section
\ref{Spinor Notation and $AdS_5$ Gravitational Field}).
An infinitesimal higher spin gauge transformation is
\be
\label{gotr}
\delta^g \go =  D\gee\,,
\ee
where $\gee (x)$ is an arbitrary infinitesimal symmetry parameter
taking values in $g$,
\be
\label{Dc}
D f = d f + [\go  \,, f ]_*\,
\ee
and
\be
\label{[]*}
[a\,,b ]_*  =a*b - b*a \,.
\ee
The higher spin curvature has the standard
homogeneous transformation law
\be
\label{Rtr}
\delta^g R =  [R  \,, \gee ]_*\,.
\ee

The higher spin equations of motion will be formulated
in terms of the higher spin curvatures and therefore admit a
zero-curvature vacuum solution with $R=0$. Since the space-time
symmetry algebra $h$ is assumed to  belong to $g$, a possible
ansatz is with all vacuum gauge fields vanishing except for
$\go_0$ taking values in $h$
\be
\label{l}
\go_0 = w_0^{AB} t_{AB} =
 \go_0^{L\,ab} L_{ab} + \lambda h_a P^a\,.
\ee
Provided that $h^a$ is nondegenerate,
the zero-curvature condition
\be
\label{Rvac}
R(\go_0 )=
(d\go_0^{AB} + \go_0^{A}{}_C \wedge \go_0^{CB})t_{AB} =0
\ee
implies that  $\go_0^{L\,ab}$ and $h^a$ identify with the gravitational
fields of $AdS_d$. Let us note that throughout this paper we  use
notation  $\go_0^{L\,ab}$ and $h^a$ for the background $AdS$ fields
satisfying (\ref{l}) but $\go^{L\,ab}$ and $e^a$ for the dynamical
gravitational fields.

Suppose there is  a theory invariant  under
the gauge transformations (\ref{gotr}).
Global symmetry is the part of the gauge transformations
that leaves invariant the vacuum solution $\go_0$. The global symmetry
parameters therefore satisfy
\be
\label{vep}
0 =  D_0 \gee^{gl} \,,
\ee
where
\be
\label{D0c}
D_0 f = d f + [\go_0  \,, f ]_*\,.
\ee
The vacuum zero-curvature equation (\ref{Rvac}) guarantees that (\ref{vep})
is formally consistent. Fixing a value of $\gee^{gl}  (x_0 )$
at some point $x_0$, (\ref{vep})  allows one to reconstruct
$\gee^{gl} (x)$ in some neighbourhood of $x_0$. Since
$D_0$ is a derivation, the star commutator of any two solutions of
(\ref{vep}) gives again some its solution. The global symmetry  algebra
therefore coincides with the algebra of star commutators at any fixed
space-time point $x_0$, which is $g$.
An important comment is that this conclusion remains true also in case
 the theory is invariant under a deformed gauge transformation
of the form
\be
\label{godeftr}
\delta \go =  D\gee  +\Delta (R,\gee )\,,
\ee
where $\Delta (R, \gee )$ denotes some $R$-dependent terms, i.e.
$\Delta (0, \gee ) =0\,.$
Indeed, all additional terms do not contribute to the
invariance condition (\ref{vep}) once the vacuum solution satisfies
(\ref{Rvac}).
In fact, as is clear from the discussion in the
section \ref{$AdS_d$ Gravity with Compensator},
the deformation of the  gauge
transformations (\ref{godeftr}) takes place in all
theories containing gravity and, in particular, in the higher spin
gauge theories.

Let us use the  perturbation expansion with
\be
\label{go01}
\go = \go_0 +\go_1 \,,
\ee
where $\go_1$ is the dynamical (fluctuational) part of the
gauge fields of the higher spin algebra $g$. Since
$R(\go_0 )=0$ we have
\be
R=R_1 +R_2\,,
\ee
where
\be
\label{R11}
R_1 = d\go_1 +\go_0 *\wedge \go_1 +  \go_1*\wedge \go_0\,,\qquad
R_2 = \go_1 * \wedge  \go_1\,.
\ee
The Abelian lowest order part of the transformation (\ref{gotr})
(equivalently, (\ref{godeftr})) has the form
\be
\label{lintr}
\delta_0 \go_1 =  D_0 \gee \,.
\ee
{}From (\ref{Rtr}) and (\ref{Rvac}) it follows that
\be
\label{lrtr}
\delta_0 R_1 = 0\,.
\ee

The idea is to construct the higher spin action from the higher spin
curvatures $R$ in the form analogous to
the gravity action (\ref{gact})
\be
\label{S3}
S = \int  U_{\Omega \Lambda} \wedge R^\Omega \wedge
R^\Lambda \,
\ee
with some $(d-4)$ - form coefficients $U_{AB}$ built from the
 frame field and compensator. ($\Omega,\Lambda $ label the adjoint
representation of $g$). To clarify whether this is
possible or not, one has to check first of all if it is true
for the free field action, i.e. whether some action of the form
\be
\label{S2}
S_2^s = \int U_{0\,\Omega \Lambda}^s \wedge R_1^\Omega \wedge R_1^\Lambda
\ee
describes the free field
dynamics of a field of a given spin $s$.
As long as $g$ is not known, a form of $R_1$ has itself
to be fixed from this requirement. In fact, the form of $R_1$ provides
an important information on the structure of $g$ fixing a pattern of the
decomposition of $g$ under the adjoint action of $h \subset g$ (up to a
multiplicity of the representations associated with a given spin $s$:
it is not {\it a priori}  known how many fields of a given spin are present in
a full higher spin multiplet). For the totally symmetric
higher spin gauge fields described by the action (\ref{fract})
this problem was solved first for case $d=4$ \cite{Fort1} and
then for any dimension both for bosons \cite{LV} and for fermions
\cite{vf}. The results of \cite{LV} are summarized in the
section \ref{Totally Symmetric Bosonic Massless Fields in any Dimension}.

As a result of (\ref{lrtr}) any action of the form (\ref{S2}) is invariant
under the Abelian free field higher spin gauge
transformations (\ref{lintr}). However, for  generic
coefficients, it not necessarily describes a consistent higher
spin dynamics. As this point is of the key importance for the analysis
of the higher spin dynamics let us explain it in somewhat more detail.
The set of 1-forms contained in $\go$ decomposes into subsets $\go^s_n$
associated with a given spin $s$. The label $n$ enumerates different
subsets associated with the same spin. (For the case $d=4$
$s$ is indeed a single number while for generic fields in
higher dimensions $s$ becomes
a vector associated with the appropriate weight vector of the $AdS_d$
algebra $o(d-1,2)$). Any subset $\go^s_n$ forms a representation of the
space-time subalgebra $h\subset g$. It further decomposes into representations
of the Lorentz subalgebra of $h$, denoted $\go_n^{s,t}$. For the case
of totally symmetric representations discussed in the Introduction,
there is a single integer parameter $ t=0,1
\ldots s-1$ that distinguishes between different Lorentz components
(for definiteness we focus here on the bosonic case of integer spins
studied in this paper). True higher spin field identifies with
$\go_n^{s,0}$.  It is called dynamical higher spin field. The rest of
the fields $\go_n^{s,t}$ with $t>0$ express in terms of
(derivatives of) the dynamical ones by
virtue of certain constraints. At the linearized level, the gauge invariant
constraints can be chosen in the form of some linear combinations of the
linearized higher spin curvatures
\be
\label{congen} \Phi^l  (R_1 ) =0
\ee
with the coefficients built from the background frame field.
By virtue of these constraints all fields $\go_n^{s,t}$
turn out to be expressed
via $t^{th}$ space-time derivatives of the dynamical field
\be
\label{hder}
\go_n^{s,t}  \sim
\Big ( \f{\p}{\lambda \p x} \Big )^t\, \left (\go_n^{s,0} \right ) +
\quad\mbox{pure gauge terms (\ref{lintr})}\,.
\ee
These expressions contain explicitly the dependence on the $AdS_d$ radius
$\lambda^{-1}$ as a result of the definition of the frame
field (\ref{defh}).

A particular example is provided with the spin 2.
Here $\go^{2,0}$ identifies with the frame field while $\go^{2,1}$
is the Lorentz connection. (We skip the label $n$ focusing on a
particular spin 2 field). The constraint (\ref{congen}) is the
linearized zero-torsion condition.

For $s>2$ the fields
$\go_n^{s,t}$ with $t\geq 2$ appear, containing second and higher
derivatives of the dynamical field. These are called extra fields.
{}From this perspective, the requirement that the free
action contains at most two space-time derivatives of the dynamical
field is equivalent to the condition that the variation
of the free action with respect to all extra fields is
identically zero
\be
\label{extvar}
\f{\delta S_2^s}{\delta \go_n^{s,t} } \equiv 0 \qquad t\geq 2\,.
\ee
It turns out that this {\it extra field decoupling
condition} fixes a form of the free action
(i.e. of $U_{0\,\Omega \Lambda}^s$)
uniquely modulo total derivatives and an overall ($s$- and $n$-
dependent) factor.
The Lorentz-type
fields  $\go_n^{s,1}$ are auxiliary, i.e. they do contribute
into the free action but express via the derivatives of the
dynamical field by virtue of their field equations equivalent to
some of the constraints (\ref{congen}).

Once the extra fields are expressed in terms of the derivatives
of the dynamical fields, the higher spin transformation law
(\ref{gotr}) (and its possible deformation (\ref{godeftr}))
describes the transformations of the
dynamical fields via their higher derivatives. Since $t$
ranges from 0 to $s-1$  one finds that the
higher spin is the higher derivatives appear in the
transformation law. Note that this conclusion is
in agreement with the general analysis of the structure
of the higher spin interactions  \cite{pos} and
conserved higher spin currents \cite{cur,gol} containing
higher space-time derivatives.

As a first step towards the non-linear higher spin dynamics one can
try the action (\ref{S3})
with $U_{\Omega \Lambda}$ proportional to
the coefficients $U_{0\,\Omega \Lambda}^s$ in the subsector of
each field of spin $s$.
This action is not invariant under the original higher spin
gauge transformations (\ref{gotr}) since $U_{\Omega \Lambda}$
cannot be an invariant tensor of $g$. Indeed, since the action is
built in terms of  differential forms without Hodge star operation,
its generic variation is
\be
\label{dS}
\delta S = -2\int  D (U_{\Omega \Lambda}) \wedge \delta \go^\Omega \wedge
R^\Lambda \,.
\ee
If $U_{\Omega \Lambda}$ would be a $g$--invariant tensor,
$S$ would be a topological invariant.
This cannot be true since the linearized action (\ref{S2}) is supposed
to give rise to nontrivial equations of motion. Therefore,
 $D (U_{\Omega \Lambda})\neq 0$. The trick is that for some
particular choice of $U_{\Omega \Lambda}$
there exists such a deformation of the gauge transformations
(\ref{godeftr}) that the action remains invariant at least in the
lowest nontrivial order, i.e. the $\go^2_1\gee$ type terms can be proved
to vanish in the variation. (Note that this is just the order at
which the difficulties with the higher-spin-gravitational interactions
were originally found \cite{diff}). In particular, we show in the
section \ref{Higher Spin Action} that this is true for the
$\N =0$ $5d$ higher spin theory. This deformation of the gauge
transformations is analogous to that resulting via
(\ref{leftsym}) from the particular
gauge fixing of the compensator field $V^A$ in the case of gravity
which, in turn, is described by the spin 2 part of the action
(\ref{S3}) equivalent to (\ref{gact}). A complication of the
Lagrangian formulation of the higher spin dynamics is that no
full-scale extension of the compensator $V^A$ to some representation
of $g$ is yet known. The clarification of this issue
is  one of the key problems on the way towards the full
Lagrangian formulation of the higher spin theory. Note, that the
full formulation of the on-mass-shell $4d$ higher spin dynamics
\cite{Ann,more} was achieved by virtue of introducing additional
compensator-type pure  gauge variables \cite{more,gol}.

Since the extra fields do
contribute into the nonlinear action it is necessary
to express them in terms of the dynamical higher spin fields to
make the nonlinear action (\ref{S3}) meaningful.
The expressions (\ref{hder}) that follow from the constraints
for extra fields
effectively induce higher derivatives into the higher spin interactions.
The same mechanism induces the negative powers of $\lambda$, the
square root of the cosmological constant, into the higher spin
interactions with higher derivatives\footnote{Note
that one can rescale the fields in such a way that
the corresponding expression (\ref{hder}) will not contain
negative powers of $\lambda$
explicitly. However, as a result of such a rescaling, $\lambda$  will
appear both in positive and in negative powers in the  structure
coefficients of the algebra $g$ and, therefore, in
the nonlinear action. From this perspective, the appearance of the
extra fields for higher spins
makes difference compared to the case of pure gravity
that allows the In\"onu-Wigner flat contraction.}.
A specific form of the
constraints (\ref{congen}) plays a crucial role
in the proof of the invariance of the action.

The program sketched in this section
was  accomplished for the $4d$ case. The free higher spin
actions of the form (\ref{S2}) were built in \cite{Fort1}. The
$4d$ higher spin algebra $g$ was then found in \cite{FVa}. In
\cite{FV1} the action (\ref{S3})
was found that described properly some cubic higher spin
interactions including the gravitational interaction.
In this paper we extend these results to
the bosonic $\N=0$ $5d$ higher spin gauge theory.

\subsection{Symmetric Bosonic Massless Fields in $AdS_d$}
\label{Totally Symmetric Bosonic Massless Fields in any Dimension}

In this section  the results of \cite{LV} are reformulated
in terms of the compensator formalism.
According to \cite{LV}, a
totally symmetric massless field of spin $s$ is described by
a collection of 1-forms
$dx^\un \go_\un{}^{a_1 \ldots a_{s-1}, b_1\ldots b_t }$ which are
symmetric in the Lorentz vector indices
$a_i$ and $b_j$ separately ($a , b \ldots  0\div d-1$),
satisfy the antisymmetry relation
\be
\label{asym}
\go_\un{}^{a_1 \ldots a_{s-1}, a_s b_2\ldots b_t }   =0 \,,
\ee
implying that symmetrization over any $s$ fiber indices gives zero,
and are traceless with respect to the fiber indices
\be
\label{tr}
\go_\un{}^{a_1 \ldots a_{s-3}c}{}_c{}^{, b_1\ldots b_t } =0 \,.
\ee
(From this condition it follows by virtue of
(\ref{asym}) that all other traces of the fiber indices are also zero).

The higher spin gauge fields associated with the spin $s$ massless
field therefore take values in the direct sum of all irreducible
representations of the $d$-dimensional massless Lorentz group
$o(d-1,1)$ described by the Young diagrams with at most two rows
such that the longest row has length $s-1$

\bigskip
\vskip -5mm
\be
\label{dia}
\begin{picture}(20,50)
\put(20,45){s-1}
\put(33,35){\circle*{2}}
\put(25,35){\circle*{2}}
\put(17,35){\circle*{2}}
\put(25,25){\circle*{2}}
\put(17,25){\circle*{2}}
\put(33,25){\circle*{2}}
\put(00,40){\line(1,0){70}}
\put(00,30){\line(1,0){70}}
\put(50,30){\line(0,1){10}}
\put(60,30){\line(0,1){10}}
\put(70,30){\line(0,1){10}}
\put(00,20){\line(1,0){60}}
\put(00,20){\line(0,1){20}}
\put(10,20){\line(0,1){20}}
\put(40,20){\line(0,1){20}}
\put(60,20){\line(0,1){20}}
\put(50,20){\line(0,1){20}}
\put(20,10){t}
\end{picture}
\ee
\vskip -10mm
\vskip 8mm
\noindent
$\go_\un{}^{a_1 \ldots a_{s-1}}$
is treated as the dynamical spin $s$
field analogous to the gravitational frame (spin 2).
The fields corresponding to the
representations with nonzero second row ($t>0$) are
auxiliary ($t=1$) or ``extra" ($t>1)$, i.e. express via derivatives
of the dynamical field by virtue of certain constraints
analogously to the Lorentz connection in the spin 2 case.
Analogously to the relationship between metric
and frame formulations of the linearized gravity,
the totally symmetric double traceless higher
spin fields used to describe the higher spin dynamics in the metric-type
formalism \cite{Fr,WF} identify with the symmetrized part of the field
$\go_\un{}^{a_1 \ldots a_{s-1}}$
\be
\varphi_{a_1 \ldots a_s} = \go_{\{ a_1\, \ldots a_s\} }\,.
\ee
The antisymmetric part in
$\go_\un{\,}^{a_1 \ldots a_{s-1}}$ can be gauge fixed to zero with the aid of
the generalized higher spin Lorentz symmetries with the parameter
$\gep{}^{a_1 \ldots a_{s-1},  b }$. That
$\varphi_{a_1 \ldots a_s}$ is double traceless is a trivial consequence of
(\ref{tr}).

The collection of the higher spin  1-forms
$dx^\un \go_\un{}^{a_1 \ldots a_{s-1}, b_1\ldots b_t }$
with all $0\leq t \leq s-1$ can be
interpreted as a result of the ``dimensional reduction''
of a 1-form $dx^\un \go_\un{}^{A_1 \ldots A_{s-1}, B_1\ldots B_{s-1} }$
carrying the irreducible representation of the $AdS_d$ algebra $o(d-1,2)$
described by the traceless two-row rectangular Young diagram of
length $s-1$
\be
\label{irre}
\go^{\{A_1 \ldots A_{s-1},A_s\} B_2\ldots B_{s-1} } =0\,,\qquad
\go^{A_1 \ldots A_{s-3}C}{}_{C,}{}^{B_1\ldots B_{s-1} } =0\,.
\ee

The linearized higher spin curvature $R_1$ has the following
simple form
\bee
\label{R1A}
R_1^{A_1 \ldots A_{s-1}, B_1\ldots B_{s-1} } &=& D_0
(\go^{A_1 \ldots A_{s-1}, B_1\ldots B_{s-1}}) =
d \go_1^{A_1 \ldots A_{s-1}, B_1\ldots B_{s-1} }\nn\\
 &{}&\ls\ls\ls\ls\ls\ls\ls\ls\ls +(s-1)\Big(
\go_0^{\{A_1}{}_{C}\wedge
\go_1^{C A_2 \ldots A_{s-1}\}, B_1\ldots B_{s-1} }
+\go_0^{\{B_1}{}_{C}\wedge
\go_1^{ A_1 \ldots A_{s-1}, C B_2\ldots B_{s-1}\} }\Big ),
\eee
where $\go_0^{AB}$ is the background $AdS_d$ gauge field
satisfying the zero curvature condition (\ref{Rvac}).

In these terms, the Lorentz covariant irreducible fields
$dx^\un \go_\un{}^{a_1 \ldots a_{s-1}, b_1\ldots b_t }$
identify with those components of
$dx^\un \go_\un{}^{A_1 \ldots A_{s-1}, B_1\ldots B_{s-1} }$
that are parallel to $V^A$ in $s-t-1$  indices and transversal in
the rest ones.
The expressions for the Lorentz components of the
linearized curvatures have the following structure
\bee
\label{R1}
\ls R_1^{a_1 \ldots a_{s-1}, b_1\ldots b_t } =
D^L \go_1^{a_1 \ldots a_{s-1}, b_1\ldots b_t }
+\tau_- (\go )^{a_1 \ldots a_{s-1}, b_1\ldots b_t }
+\tau_+ (\go )^{a_1 \ldots a_{s-1}, b_1\ldots b_t }
\eee
with
\be
\label{t-}
\tau_- (\go )^{a_1 \ldots a_{s-1}, b_1\ldots b_t }=
\alpha h_c\wedge \go_1^{a_1 \ldots a_{s-1}, b_1\ldots b_t c} \,,
\ee
\be
\label{t+}
\tau_+ (\go )^{a_1 \ldots a_{s-1}, b_1\ldots b_t }=
\beta \Pi \left(h^{b_1}\wedge
\go_1^{a_1 \ldots a_{s-1}, b_2\ldots b_{t}}\right )\,,
\ee
where $D^L$ is the background Lorentz covariant differential,
$\Pi$ is the projection operator to the irreducible representation
described by the traceless Young diagram of the Lorentz algebra
$o(d-1,1)$ with $s-1$ and $t$  cells
in the first and second rows, respectively,
and $\alpha$ and $\beta$ are some coefficients depending on
$s$, $t$ and $d$ and fixed in such a way that
\be
\label{tau+-}
(\tau_- )^2 =0\,,\qquad (\tau_+ )^2 =0\,,\qquad
(D^L )^2 + \{\tau_- , \tau_+ \} =0 \,.
\ee
For the explicit expressions of $\ga$, $\gb$ and $\Pi$
we refer the reader to the original paper \cite{LV}.
The explicit spinor version of the formula (\ref{R1}) for $d=5$
will be given in section \ref{d5 Higher Spin Gauge Fields}.

The quadratic action functional for the massless spin $s$
field equivalent to the  $AdS_d$  deformation
of the action (\ref{fract}) has the following simple form \cite{LV}
\footnote{In this paper we use the normalization of fields in terms
of the AdS parameter $\lambda$ different from that of \cite{LV}.
Namely, we assume that $\lambda$
enters only via the definition of the frame field $E^A$ while in \cite{LV}
the fields were normalized in such a way that their expressions in terms
of the derivatives of the dynamical higher spin field were free from negative
powers of $\lambda$. This difference results
in the different form of the  dependence of
the higher spin action on $\lambda$.}
\bee
\label{S2S}
 S_2^s &=& \half \chi(s) \int_{M^d} \sum_{p=0}^{s-2}
\frac{[(p+1)!]^2}{(d+p-3)!}
\varepsilon_{c_1 \ldots c_d}
h^{c_5}\wedge \ldots \wedge h^{c_d}\nn\\
  &\wedge&
R_1^{c_1 a_1 \ldots a_{s-2}, c_3 b_1\ldots b_{p} }\wedge
R_1^{c_2}{}_{a_1 \ldots a_{s-2},}{}^{ c_4}{}_{ b_1\ldots b_{p }}\,.
\eee
It is fixed up to an overall normalization
factor $\chi (s)$ by the conditions that it is $P$-even
and its variation with respect to the ``extra fields''
is identically zero,
\be
\label{cond}
\frac{\delta S^s_2}{\delta \go_\un{}^{a_1 \ldots a_{s-1}, b_1\ldots b_t }}
\equiv 0 \qquad \mbox{for}  \quad t\geq 2\,.
\ee

Let us explain how one can derive a $o(d-1,2)$
covariant form of the same action with the aid of the compensator $V^A$.
Taking into account the irreducibility properties (\ref{irre})
one finds that the general form of the $P$-even action written
in terms of differential forms is
\bee
\label{gcovdact}
S^s_2&=&\half
\int_{M^d}\sum_{p=0}^{s-2}a (s,p)
\gep_{A_1 \ldots A_{d+1}}h^{A_5}\wedge\ldots     \wedge h^{A_{d}}
V^{A_{d+1}}
    V_{C_1}\ldots V_{C_{2(s-2-p)}}\nn\\
&{}&\ls\ls\ls R_1^{A_1 B_1 \ldots B_{s-2}}{}_,{}^{A_2 C_1 \ldots C_{s-2-p}
D_1\ldots D_p}\wedge R_1^{A_3}{}_{B_1 \ldots B_{s-2},}{}^{ A_4 C_{s-1-p} \ldots
C_{2(s-2-p)}}{}_{D_1\ldots  D_p }\,.
\eee
Consider a general variation of $S^s_2$ with respect to
$\go^{A_1 \ldots A_{s-1}}{}_,{}^{ B_1\ldots B_{s-1} }$. Using
that
\be
\delta R_1^{A_1 \ldots A_{s-1}}{}_,{}^{ B_1\ldots B_{s-1} }=
D_0 \delta \go_1^{A_1 \ldots A_{s-1}}{}_,{}^{ B_1\ldots B_{s-1} }\,,
\ee
where $D_0$ is the background derivative, one integrates by parts
taking into account that  $D_0 (V^A ) =h^A$, $D_0 (h^A ) =0$.
With  the help of the irreducibility conditions (\ref{irre}),
the identity (\ref{eid}) and the identity
\bee
\gep_{A_1 \ldots A_5 B_6 \ldots  B_{d+1}}\!\!\!&{}&\!\!\!\!\!h^C \wedge
h^{B_6}\wedge\ldots     \wedge h^{B_{d+1}}\nn\\
&&\ls\ls\ls\ls\ls\ls=(d-3)^{-1}
\Big (
\gep_{A_1 A_2 A_3 A_4 B_5 \ldots  B_{d+1}}\, \delta_{A_5}^C-
\gep_{A_1 A_2 A_3 A_5 B_5 \ldots  B_{d+1}}\, \delta_{A_4}^C\,
+
\gep_{A_1 A_2 A_4 A_5 B_5 \ldots  B_{d+1}}\, \delta_{A_3}^C \,
\nn\\
                 &{}&\!\!\ls\ls\ls\ls -
\gep_{A_1 A_3 A_4 A_5 B_5 \ldots  B_{d+1}} \,\delta_{A_2}^C \,
+
\gep_{A_2 A_3 A_4 A_5 B_5 \ldots  B_{d+1}} \,\delta_{A_1}^C \Big )
h^{B_5}\wedge\ldots     \wedge h^{B_{d+1}}\,,
\eee
which expresses the simple fact that the total antisymmetrization
of any set of $d+2$ vector indices $A_i $ is zero, one finds
\bee
\label{covvar}
\delta     S_2^s\!\!\!&=& \!\!\!\!- \frac{\lambda}{d-3}
\int_{M^d}\sum_{p=0}^{s-2} \Big (\frac{(s-p)(d-7 +2 (s -p))}{s-p-1}
a (s,p) - (s-p-1) a (s,p-1 ) \Big )\nn\\
\ls\ls&{}&V_{C_1} \ldots V_{C_{2(s-p)-3}}
\gep_{A_1 \ldots A_{d+1}} V^{A_4}
\wedge h^{A_5}\wedge\ldots     \wedge h^{A_{d+1}}\wedge \nn\\
\ls\ls&{}&
\Big (\delta \go_1^{A_1 B_1 \ldots B_{s-2}}{}_,{}^{A_2 C_1 \ldots C_{s-2-p}
D_1\ldots D_p}\wedge R_1^{A_3}{}_{B_1 \ldots B_{s-2},}{}^{ C_{s-1-p} \ldots
C_{2(s-p)-3}}{}_{D_1\ldots  D_p }  \nn\\
\ls\ls&{}&
+R_1^{A_1 B_1 \ldots B_{s-2}}{}_,{}^{A_2 C_1 \ldots C_{s-2-p}
D_1\ldots D_p}\wedge
\delta \go_1^{A_3}{}_{B_1 \ldots B_{s-2},}{}^{ C_{s-1-p} \ldots
C_{2(s-p-3)}}{}_{D_1\ldots  D_p } \Big )\,.\nn\\
\eee
The idea is to require all the terms in (\ref{covvar}) to vanish
except for the term at $p=0$. This condition fixes the coefficients
$a (s,p)$ up to a normalization factor $\tilde{a} (s)$ in the form
\be
\label{al}
a (s,p) = -\tilde {a} (s) \lambda^{-1} (d-3)
\frac{(d-5 +2 (s-p-2))!!\, (s-p-1)(s-2)!}{s\,(d-3 +2 (s-2))!!  \,(s-p-2)!}\,.
\ee
As a result the variation (\ref{covvar}) acquires the form
\bee
\label{dynvar}
\delta     S_2^s&=& \!\!\tilde{a} (s)
\int_{M^d}
V_{C_1} \ldots V_{C_{2s-3}}
\gep_{A_1 \ldots A_{d+1}} V^{A_4}
h^{A_5}\wedge\ldots\wedge h^{A_{d+1}}\wedge\nn\\
&{}&
\Big (\delta \go_1^{A_1 B_1 \ldots B_{s-2}}{}_,{}^{A_2 C_1 \ldots C_{s-2}}
\wedge R_1^{A_3}{}_{B_1 \ldots B_{s-2},}{}^{ C_{s-1} \ldots
C_{2s-3}}{}  \nn\\
&{}&
+R_1^{A_1 B_1 \ldots B_{s-2}}{}_,{}^{A_2 C_1 \ldots C_{s-2}}\wedge
\delta \go_1^{A_3}{}_{B_1 \ldots B_{s-2},}{}^{ C_{s-1} \ldots
C_{2s-3}}{} \Big )\,.
\eee
This formula implies that the free action (\ref{gcovdact}),
(\ref{al}) essentially depends only on
the $V^A$-transversal
parts of
\be
\label{cphys}
\go_\un{}_{A_1 \ldots A_{s-1} }=
\go_\un{}_{A_1 \ldots A_{s-1} }{}_{,B_1\ldots B_{s-1}} V^{B_1} \ldots
V^{B_{s-1}}
\ee
and
\be
\label{clor}
\go_\un{}_{A_1 \ldots A_{s-1} }{}_{,B_1}=
\go_\un{}_{A_1 \ldots A_{s-1} }{}_{,B_1\ldots B_{s-1}} V^{B_2} \ldots
V^{B_{s-1}}\,.
\ee
These fields identify respectively with
the frame--like dynamical higher spin field
$\go_\un{}^{a_1 \ldots a_{s-1} }$  and the Lorentz connection--like
auxiliary field $\go_\un{}^{a_1 \ldots a_{s-1}, b }$ expressed
in terms of the first derivatives of the frame-like field by virtue of
its equation of motion equivalent to the ``zero torsion condition"
\be
\label{ctor}
0= T_1 {}_{A_1 \ldots A_{s-1} }\equiv
R_{1\,A_1 \ldots A_{s-1} }{}_{,B_1\ldots B_{s-1}} V^{B_1} \ldots
V^{B_{s-1}}\,.
\ee
Insertion of the expression for
$\go_\un{}^{a_1 \ldots a_{s-1}, b }$ into (\ref{gcovdact})
gives rise to the higher spin action expressed entirely
(modulo total derivatives) in terms
of $\go_\un{}^{a_1 \ldots a_{s-1} }$ and its first derivatives.
Since the linearized curvatures (\ref{R1})
are by construction invariant under the Abelian higher spin gauge
transformations
\bee
\label{dgo}
\delta \go_1{}^{A_1 \ldots A_{s-1}, B_1\ldots B_{s-1} }  =
D_0 \e^{A_1 \ldots A_{s-1}, B_1\ldots B_{s-1} }
\eee
with the higher spin gauge parameters
$\gee^{A_1 \ldots A_{s-1}, B_1\ldots B_{s-1} }$,
the resulting action possesses required
higher spin gauge symmetries and therefore describes
correctly the free field higher spin dynamics in $AdS_d$.
In particular, the generalized Lorentz-like transformations
with the gauge parameter
\be
\gep_{A_1 \ldots A_{s-1} }{}_{,B_1} (x)
\ee
guarantee that only the totally symmetric part of the gauge
field  (\ref{cphys})  equivalent to $\varphi_{m_1\ldots m_s}$
contributes to the action.
Analogously, the auxiliary Lorentz-type higher spin field
has pure gauge components associated with the
generalized Lorentz-type transformations
parameter  described by the two-row Young diagram
with two cells in the second row. These components do
not express in terms of the dynamical higher spin field.
However, the invariance
with respect to the gauge transformations (\ref{dgo})
guarantees that these pure gauge components do not contribute
into the action.

Although the extra fields
$\go_\un{}^{a_1 \ldots a_{s-1}, b_1\ldots b_t }$ with $t\geq 2$
do not contribute to the free action, as we have learned from
the four-dimensional case \cite{FV1} they do contribute at the
interaction level. To make such interactions meaningful, one has to
express the extra fields in terms of the dynamical ones modulo
pure gauge degrees of freedom.
This is achieved  by imposing constraints \cite{LV}
\be
\label{conv}
\gep^{a_1 b_1}{}_{e_1\ldots e_{d-4} cf}
h^{e_1}\wedge \ldots \wedge h^{e_{d-4}} \wedge
\tau_+(R_1 ){}^{c a_2 \ldots a_{s-1},f b_2\ldots b_t }  =0
\ee
(total symmetrizations within the groups of indices
$a_i$ and $b_j$ is assumed). The covariant version of these constraints
is
\be
\label{cconv}
\gep^{A_1 B_1}{}_{E_1\ldots E_{d-4} CFG} V^G
h^{E_1}\wedge \ldots \wedge h^{E_{d-4}} \wedge
\tau_+(R_1 ){}^{C A_2 \ldots A_{s-1},F B_2\ldots B_{s-1} }  =0\,.
\ee
The covariant expressions for the operators $\tau_\pm$
are complicated and will not be given here for general $d$.
For $d=5$ they are given in section
\ref{d5 Higher Spin Gauge Fields} in the spinor formalism.

An important fact is \cite{LV} that,
by virtue of these constraints, most of
the higher spin field strengths vanish on-mass-shell according to the
following relationship
referred to as the First On-Mass-Shell Theorem
\bee
\label{CMT}
R_1^{a_1 \ldots a_{s-1}, b_1\ldots b_t }\!\!&=&\!\!\!
X^{a_1 \ldots a_{s-1}, b_1\ldots b_t }
(\frac{\delta{S_2}}{\delta \go_{dyn}})
\qquad for \quad t < s-1\,,\nn\\
R_1^{a_1 \ldots a_{s-1}, b_1\ldots b_{s-1} }\!\!&=&\!\!\!h_{a_s} \wedge h_{b_s}
C^{a_1 \ldots a_{s}, b_1\ldots b_{s} }
+
X^{a_1 \ldots a_{s-1}, b_1\ldots b_{s-1} }
(\frac{\delta{S_2}}{\delta \go_{dyn}})\,.
\eee
Here
$
X^{a_1 \ldots a_{s-1}, b_1\ldots b_t }
(\frac{\delta{S^s_2}}{\delta \go_{dyn}})
$
are some linear functionals of the left-hand-sides of the
free field equations
$\frac{\delta{S^s_2}}{\delta \go_{dyn}}=0$ for the spin $s$
dynamical one-forms $\go_{dyn}^{a_1 \ldots a_{s-1}}$.
The 0-forms $C^{a_1 \ldots a_{s}, b_1\ldots b_{s} }$
are described by the traceless two-row rectangular
Young diagrams of length $s$ and parametrize
those components of the higher spin field strengths that can remain
nonvanishing when the field equations and constraints are
satisfied. These generalize the Weyl tensor in gravity ($s=2$) that
parametrizes the components of the Riemann tensor
allowed to be nonvanishing when  the zero-torsion
constraint and Einstein equations (requiring the Ricci tensor to
vanish) are imposed. The covariant version of (\ref{CMT}) is
\be
\label{ccomt}
R_1^{A_1 \ldots A_{s-1}, B_1\ldots B_{s-1} }=h_{A_s} \wedge h_{B_s}
C^{A_1 \ldots A_{s}, B_1\ldots B_{s} }\,
+ X^{A_1 \ldots A_{s-1}, B_1\ldots B_{s-1} }
(\frac{\delta{S^s_2}}{\delta \go_{dyn}})
\ee
with
$C^{A_1 \ldots A_{s}, B_1\ldots B_{s} }$
described by the traceless $V^A$--transversal two-row rectangular
Young diagram of length $s$, i.e.
\be
C^{\{A_1 \ldots A_{s},A_{s+1}\} B_2\ldots B_{s} }\, =0\,,\qquad
\ee
\be
C^{A_1 \ldots A_{s-2}CD, B_1\ldots B_{s} }\,\eta_{CD} =0\,,\qquad
C^{A_1 \ldots A_{s-1}C, B_1\ldots B_{s} }\,V_{C} =0\,.
\ee

For completeness, let us present the unfolded equations of motion
for all free integer spin massless higher spin fields
in $AdS_d$ corresponding
to the totally symmetric representations of the Wigner little group
(more precisely, totally symmetric lowest weight vacua of the
irreducible representations of the $AdS_d$ algebra $o(d-1,2)$).
The content of the Central On-Mas-Shell Theorem is that the
equations of motion for massless free fields of all spins can be
written in the form
\be
\label{cmt1}
R_1^{A_1 \ldots A_{s-1}, B_1\ldots B_{s-1} }=h_{A_s} \wedge h_{B_s}
C^{A_1 \ldots A_{s}, B_1\ldots B_{s} }\,,
\ee
\be
\label{cmt2}
D_0
C^{A_1 \ldots A_{u}, B_1\ldots B_{s} } =0\q u\geq s\,,
\ee
where
\be
D_0 = D_0^L +\sigma_- + \sigma_+ \,,
\ee
$D_0^L$ is the vacuum Lorentz covariant derivation and
the operators $\sigma_\pm$ have the form
\be
\sigma_- (C)^{A_1 \ldots A_{u}, B_1\ldots B_{s} }
=(u-s+2) E_C C^{A_1 \ldots A_{u}C, B_1\ldots B_{s} }
+s E_C C^{A_1 \ldots A_{u}B_s, B_1\ldots B_{s-1} C }
\ee
\bee
\sigma_+ (C)^{A_1 \ldots A_{u}, B_1\ldots B_{s} }
&=&u\lambda^2 \Big (
\f{d+u+s-4}{d+2u -2}E^{A_1} C^{A_2 \ldots A_{u}, B_1\ldots B_{s} }\nn\\
&-&\f{s}{d+2u-2}
\eta^{A_1 B_1} E_C C^{A_2 \ldots A_{u},C B_2\ldots B_{s} }\nn\\
&-&\f{(u-1)(d+u+s-4)}{(d+2u-2)(d+2u-4)}
\eta^{A_1 A_2} E_C C^{A_3 \ldots A_{u}C, B_1\ldots B_{s} }\nn\\
&+&\f{s(u-1)}{(d+2u-2)(d+2u-4)}
\eta^{A_1 A_2} E_C C^{A_3 \ldots A_{u}B_1 ,C B_2\ldots B_{s} }\Big )\,\nn\\
\eee
(total symmetrization within the groups of indices $A_i$ and $B_j$
is assumed).
The set of 0-forms $C^{A_1 \ldots A_{u}, B_1\ldots B_{s} } $
consists of all two-row traceless $V^A -$transversal
Young diagrams with the second row of length $s$, i.e.
\be
C^{\{A_1 \ldots A_{u},A_{u+1}\} B_2\ldots B_{s} }\, =0\,,\qquad
\ee
\be
C^{A_1 \ldots A_{u-2}CD, B_1\ldots B_{s} }\,\eta_{CD} =0\,,\qquad
C^{A_1 \ldots A_{u-1}C, B_1\ldots B_{s} }\,V_{C} =0\,.
\ee
The equations (\ref{cmt1}) (being a consequence of the First On-Mass-Shell
Theorem) and (\ref{cmt2})
are equivalent to the free equations of motion of
(totally symmetric) massless fields of all spins
in $AdS_d$ along with some constraints that express an infinite
set of auxiliary variables via higher derivatives of the
dynamical fields of all spins.
The proof of the Central On-Mass-Shell Theorem is analogous to
that given in the $su(2,2)$ notation in section
\ref{5d Central On-Mass-Shell Theorem}  for the $5d$ case.
The Central On-Mass-Shell Theorem plays the key role in many
respects and, in particular, for the analysis of interactions
as was originally demonstrated in \cite{Ann} where it was proved
for the $4d$ case.

Note that, as shown in \cite{SVsc},
the equations of motion of massless scalar coincide with the
sector of equations (\ref{cmt2}) with $s=0$. Analogously, the equations
(\ref{cmt2}) with $s=1$ impose the Maxwell equations on the spin 1
potential (1-form) $\go$.

\section{Compensator Formalism in $su(2,2)$ Notation}
\label{Spinor Notation and $AdS_5$ Gravitational Field}

It is well known that the
$AdS_5$ (equivalently, $4d$ conformal) algebra $o(4,2)$
is isomorphic to $su(2,2)$ and, as such, admits  realization
in terms of oscillators \cite{BG}
\be
\label{osc} [a_\ga ,b^\gb
]_*=\delta_\ga{}^\gb \,,\qquad [a_\ga ,a_\gb ]_*=0\,,
\qquad [b^\ga ,b^\gb]_*=0\,
\ee
$\ga,\gb =1\div 4$.
Here we use the star product realization of the algebra
of oscillators that describes the totally symmetric
(i.e., Weyl) ordering
\bee
\label{prod}
\!\! (f*g)(a,b)\!\!\!&=&\!\!\!\!\frac{1}{(\pi)^{8}}
\!\int\!\! d^{4}u d^{4} v d^{4}s d^{4} t
f(a\!+\!u ,b\!+\!t)g(a\!+\!s ,b\!+\!v\! )
\exp2(\!s_\ga t^\ga\!-\!u_\ga v^\ga)\nn\\
&=& e^{\half \left ( \f{\p^2}{\p s_\ga \p t^\ga} -
\f{\p^2}{\p u^\ga \p v_\ga} \right )} f (a+s , b+u ) g (a+v , b+t )
\Big|_{s=u=t=v=0}\,.
\eee
It is straightforward to see that this star product is
associative and gives rise to the commutation relations (\ref{osc})
via (\ref{[]*}).
The associative star product algebra with eight generating elements $a_\ga$
and $ b^\gb$ is called Weyl algebra $A_4$. Let us note that the star product
algebras relevant to the higher spin gauge theory (in, particular, the
one used throughout this paper) are
treated as the algebras of polynomials or formal power series thus being
different from the star product algebras of functions  regular
at infinity that are relevant to the noncommutative Yang-Mills
theory \cite{ncYM}. One important difference concerns the definitions
of the invariant trace operations because, as shown in \cite{Fort2},
the star product algebras  of formal power series possess  a uniquely
defined supertrace operation but  admits no usual trace at all (like
the one used in the non-commutative Yang-Mills theory). It is worth
to mention that the superstructure underlying the supertrace of the
polynomial star product algebras is just appropriate in the context
of the spinor interpretation of the generating elements like $a_\ga$
and $b^\gb$ in the $5d$ higher spin theory studied in this paper.

The Lie algebra $gl_4 $ is spanned by the bilinears
\be
T_\ga{}^\gb = a_\ga b^\gb \equiv \half (a_\ga * b^\gb  +b^\gb * a_\ga )\,.
\ee
The central component is associated with the generator
\be
\label{N}
N= a_\ga b^\ga \equiv \half ( a_\ga * b^\ga + b^\ga  * a_\ga )
\ee
while the traceless part
\be
t_\ga{}^\gb = a_\ga b^\gb -\frac{1}{4}\delta_\ga{}^\gb N
\ee
spans $sl_4$. The
$su(2,2)$ real form of $sl_4 (\C)$ results from
 the reality conditions
\be
\label{re}
\bar{a}_\ga  =  b^\gb C_{\gb \ga}\,,\qquad
\bar{b}^\ga =  C^{\ga\gb} a_\gb \,,
\ee
where bar denotes the complex conjugation while
$C_{\ga\gb}=-C_{\gb\ga}$ and $C^{\ga\gb}=-C^{\gb\ga}$
are some real antisymmetric matrices satisfying
\be
C_{\ga\gamma} C^{\gb\gamma} = \delta _\ga^\gb\,.
\ee

The oscillators $b^\ga$ and $a_\ga$ are in  the fundamental
 and the conjugated fundamental representations of $su(2,2)$ equivalent
to the two spinor representations of $o(4,2)$.
A $o(6)$  complex vector $V^A$ ($A=0\div 5$) is
equivalent to the antisymmetric bispinor $V^{\ga\gb}=-V^{\gb\ga}$
having six independent components (equivalently, one can use
$V_{\ga\gb} = \half \varepsilon_{\ga\gb\gga\gd}V^{\gga\gd}$ where
$\varepsilon_{\ga\gb\gamma\delta}$ is the $sl_4$ invariant totally
antisymmetric tensor ($\varepsilon_{1234} =1$)).
A  $o(4,2)$ real vector $V^A$ is described by the antisymmetric
bispinor $V^{\ga\gb}$ satisfying the reality condition
\be
\overline{V}^{ \gamma \delta}
C_{\gamma \ga} C_{\delta \gb}=
\half \varepsilon_{\ga\gb\gamma\delta}V^{\gamma\delta}\,.
\ee
One can see that the invariant norm of the vector
\be
\label{norm}
V^2 = {V}_{\ga\gb} V^{\ga\gb}
\ee
has the signature $(++----)$. The vectors with $V^2 >0$ are time-like
while those with $V^2 <0$ are space-like.
To perform a reduction
of the representations of the $AdS_5$ algebra $su(2,2)\sim o(4,2)$
into representations of its Lorentz subalgebra $o(4,1)$
we introduce a
$su(2,2)$ antisymmetric compensator $V^{\ga\gb}$ with positive
square (\ref{norm}), which is the spinor analog of the
compensator $V^A$ of section \ref{Generalities}.
The Lorentz algebra is identified with its
stability subalgebra. (Let us note that $V^{\ga\gb}$
must be different from the form $C^{\ga\gb}$ used in the definition of
the reality conditions (\ref{re}) since the latter is space-like
and therefore has $sp(4; R )\sim o(3,2)$ as its stability algebra.)

We shall treat $V^{\ga\gb}$ as a symplectic form that allows one to
raise and lower spinor indices in the  Lorentz covariant way
\be
A^\ga = V^{\ga\gb}A_\gb \,,\qquad
A_\ga = A^\gb V_{\gb\ga}\,,
\ee
Using that the total antisymmetrization over any four
indices is proportional to the $\varepsilon$  symbol,
we normalize $V^{\ga\gb}$ so that
\be
\label{vno1}
V_{\ga\gb}   V^{\ga\gamma} = \delta_\ga{}^\gamma \,,\qquad
V_{\ga\gb}=\half \varepsilon_{\ga\gb\gga\gd}V^{\gga\gd}\,,
\ee
\be
\varepsilon_{\ga\gb\gamma\delta} = V_{\ga\gb}V_{\gamma\delta}
+V_{\gb\gamma}V_{\ga\delta} +V_{\gamma\ga}V_{\gb\delta}\,,
\ee
\be
\varepsilon^{\ga\gb\gamma\delta} = V^{\ga\gb}V^{\gamma\delta}
+V^{\gb\gamma}V^{\ga\delta} +V^{\gamma\ga}V^{\gb\delta}\,.
\ee

In these terms the Lorentz subalgebra is spanned by the generators
symmetric in the spinor indices
\be
\label
{lor}
L_{\ga\gb} = \half (t_{\ga\gb} + t_{\gb\ga} )
\ee
while the $AdS_5$ translations are associated with the antisymmetric
traceless generators
\be
P_{\ga \, \gb} = \half (t_{\ga\gb} - t_{\gb\ga} )\,.
\ee
The gravitational fields are identified with the
gauge fields taking values in the $AdS_5$ algebra $su(2,2)$
\be
w =w{}^\ga{}_\gb a_\ga b^\gb  \,.
\ee
The invariant definitions of the frame field and Lorentz
connection for a $x-$depen\-dent compensator $V^{\ga\gb} (x)$ are
\be
\label{h}
E^{\ga\gb} =D V^{\ga\gb} \equiv  dV^{\ga\gb} +
w{}^\ga{}_\gamma V^{\gamma\gb} +
w{}^\gb{}_\gamma V^{\ga\gamma} \,,\qquad
\ee
\be
\go^{L\,\ga}{}_\gb = w{}^\ga{}_\gb +\half
E^{\ga\gamma} V_{\gamma \beta}\,.
\ee
The normalization condition (\ref{vno1}) implies
\be
\label{tr1}
E_{\ga\gb} =-D V_{\ga\gb}\,, \qquad
E_\ga{}^\ga =0\,.
\ee

The non-degeneracy condition implies
 that $ E^{\ga \gb}$ spans a basis of the
$5d$ 1-forms. The basis $p$-forms $E_p$ can be realized as
\be
E_2^{\ga\gb}=E_2^{\gb\ga} =E^{\ga}{}_\gamma \wedge E^{\gb\gamma}\,,
\ee
\be
E_3^{\ga\gb}=E_3^{\gb\ga} =E_2^{\ga}{}_\gamma \wedge E^{\gb \gamma }\,,
\ee
\be
E_4^{\ga\gb}=-E_4^{\gb\ga} =E_3^{\ga}{}_\gamma \wedge E^{\gb \gamma}\,,
\ee
\be
\label{E5}
E_5=E_4^{\ga}{}_\gamma \wedge E_{\ga}{}^{\gamma}\,.
\ee
The following useful relationships hold as a consequence
of the facts that $5d$ spinors have four components
and the frame field is traceless (\ref{tr1})
\be
\label{id1}
E^{\ga\gb}\wedge E^{\gga\gd} = \half (
V^{\ga\gga}E_2^{\gb\gd}
-V^{\gb\gga}E_2^{\ga\gd}
-V^{\ga\gd}E_2^{\gb\gga}
+V^{\gb\gd}E_2^{\ga\gga} )\,,
\ee
\be
E_2^{\ga\gb}\wedge E^{\gga\gd} = -\frac{1}{3} (
V^{\ga\gga}E_3^{\gb\gd}+
V^{\gb\gga}E_3^{\ga\gd}
-V^{\gb\gd}E_3^{\ga\gga}
-V^{\ga\gd}E_3^{\gb\gga}
+V^{\gga\gd}E_3^{\ga\gb} )\,,
\ee
\be
E_{4\ga}{}^\ga =0\,,
\ee
\be
E^{\ga\gb}\wedge E_3 ^{\gga\gd}
= -\frac{1}{4} (V^{\ga\gga}E_4^{\gb\gd}
-V^{\gb\gga}E_4^{\ga\gd}
+V^{\ga\gd}E_4^{\gb\gga}
-V^{\gb\gd}E_4^{\ga\gga} )\,,
\ee
\be
\label{id2}
E_4^{\ga\gb} \wedge E^{\gga\gd} = -\frac{1}{20}(2V^{\ga\gga} V^{\gb\gd} -
2V^{\ga\gd}V^{\gb\gga} -V^{\ga\gb}V^{\gga\gd} ) E_5 \,.
\ee

The gravitational field $w$ describes the $AdS_5$
geometry provided that $w=\go_0$ satisfies the
zero-curvature equation
\be
\label{ads5}
d\go_0 + \go_0 \wedge * \go_0  =0
\ee
and the frame 1-form is non-degenerate.
The background frame field and Lorentz connection
will be denoted $h=h^\ga{}_\gb a_\ga b^\gb$ and
$\go_0^L=\go_0^{L\,\ga}{}_\gb a_\ga b^\gb$, respectively. The vacuum values of
the $p$-forms $E_p^{\ga\gb}$ are denoted $H_p^{\ga\gb}$.

\section{${\bf su(2,2)}$ - ${\bf o(4,2)}$ Dictionary}
\label{$su(2,2)$ - $o(4,2)$ Dictionary}

To make contact between the tensor and spinor forms
of the higher spin dynamics one has to
identify in terms of $o(4,2)$ the irreducible finite-dimensional
representations of $su(2,2)$ described by
a pair of mutually conjugated traceless $su(2,2)$ multispinors
\be
\label{Xsu}
X^{\ga_1 \ldots \ga_n }{}_{\gb_1 \ldots \gb_m} \oplus
\overline{X}^{\gb_1 \ldots \gb_m }{}_{\ga_1 \ldots \ga_n} \,,\qquad
X^{\ga_1 \ldots \ga_{n-1}\gga }{}_{\gb_1 \ldots \gb_{m-1}\gga}=0\,.
\ee
The result is that for even $n+m$ the representation (\ref{Xsu}) is
equivalent to the representation of $o(4,2)$ described by the
traceless tree-row Young diagram having two rows of equal lengths
$\half |n+m|$ and the third one of length $\half |n-m|$.
In other words, the $o(4,2)$ form of the representation (\ref{Xsu})
is described by the tensor
$X_{A_1 \ldots A_p\,,B_1 \ldots B_p \,,C_1\ldots C_q }$ with
$p=\half |n+m|$, $q=\half|n-m|$, which is
separately symmetric with respect to the indices $A_i$, $B_i$ and
$C_i$ and satisfies the conditions
\be
X_{\{A_1 \ldots A_p\,,A_{p+1}\}B_2 \ldots B_p \,,C_1\ldots C_q }=0\,,\qquad
X_{A_1 \ldots A_p\,,\{B_1 \ldots B_p \,,B_{p+1}\}C_2\ldots C_q }=0\,
\ee
and
\be
\label{Xtr}
\eta^{D_1 D_2}
X_{D_1D_2 A_3 \ldots A_p\,,B_1 \ldots B_p \,,C_1\ldots C_q }=0\,.
\ee
({}From these conditions it follows that all other traces vanish as well.)
One example of this identification is provided by the isomorphism between
$X^{\ga\gb}$ (with its conjugate $\overline{X}_{\ga\gb}$)
 and the 3-form representation of $o(4,2)$
$X_{A\,,B\,,C}$ being totally antisymmetric in its indices.

For the case of half-integer spins with odd $n+m$, the identification
is analogous with the tensor-spinor
$X_{A_1 \ldots A_p\,,B_1 \ldots B_p \,,C_1\ldots C_q \,;\hat{\ga}}$
carrying the $o(4,2)$ spinor index $\hat{\ga}$,
$2p=n+m-1$, $2q=|n-m|-1$ and the $\gga -$transversality
condition with respect to all indices in addition to the tracelessness
condition (\ref{Xtr}).

A particular case of a self-conjugated traceless multispinor
\be
\label{X2}
X^{\ga_1 \ldots \ga_n }{}_{\gb_1 \ldots \gb_n}\,,\qquad
X^{\ga_1 \ldots \ga_{n-1}\gamma }{}_{\gb_1 \ldots \gb_{n-1} \gamma} =0
\ee
is most important for this paper. Such a tensor is
equivalent to the representation of $o(4,2)$ described
by a length-$n$ rectangular traceless two-row Young diagram,
i.e. to
\be
\tilde{X}_{A_1 \ldots A_n\,,B_1 \ldots B_n}\,,
\ee
which is separately symmetric in the indices $A_k$ and $B_k$,
has all traces zero and is subject to the condition that
symmetrization of any $n+1$ indices gives zero.
One way
to see this isomorphism is to compare the dimensions of the
representations to make sure that they are both equal to
$\frac{(2n+3)(n+1)^2 (n+2)^2}{12}\,.$
It is easy to see that this formula is true
from the $sl_4$ side. The computation in terms of $o(4,2)$
is more complicated. The dimensionality of the representation of
the orthogonal algebra $o(d)$ described by the two-row traceless
rectangular diagram of length $s$ is
\be
{\cal N}(s,d)=
\frac{(2s +d -2)!(s+d-4)!(s+d-5)!}{(d-2)!(d-4)! s! (s+1)!(2s +d -5)!}\,.
\ee
For $n=s$ and $d=6$ one finds the desired result.
For $n=1$ the isomorphism between the adjoint representations of
$su(2,2)$ and $o(4,2)$ is recovered. Note that the analogous analysis
of the representation (\ref{X2}) of $su(2,2)$ was done in \cite{SSd}
in terms of representations of the $5d$ Lorentz algebra
$o(4,1)\subset o(4,2)$.

In accordance with the analysis of section
\ref{Totally Symmetric Bosonic Massless Fields in any Dimension}
(and of \cite{SSd}) we conclude
that $5d$  spin $s$ bosonic gauge fields can be described by 1-forms
$\go^{\ga_1 \ldots \ga_{s-1}}{}_{\gb_1 \ldots \gb_{s-1}}$
which are traceless multispinors symmetric in the upper and
lower indices. Totally symmetric spin $s$ fermionic tensor-spinor
representations are described by the gauge fields
$\go^{\ga_1 \ldots \ga_{s-1/2}}{}_{\gb_1 \ldots \gb_{s-3/2}}$
and their conjugates.

All other representations in the set (\ref{Xsu}) do not correspond to
the sets of gauge fields associated with the totally symmetric
tensor(-spinor) fields. These are expected to underly the description
of the mixed symmetry  $AdS_5$ massless fields to be developed.
According to \cite{BMV}, such fields are inequivalent to
the totally symmetric higher spin fields in the $AdS$ regime, although
reduce in the flat limit to some combinations of the higher spin fields
associated with the totally symmetric representations of the flat Wigner
little algebra. As argued in \cite{BHS}, the fields
$\go^{\ga_1 \ldots \ga_{p}}{}_{\gb_1 \ldots \gb_{q}}$ with $|p-q|\geq 2$
necessarily appear in the $5d$ higher spin gauge theories with
$\N\geq 2$ extended supersymmetry.
This raises the important problem of the development of the
formulation of the corresponding massless fields in $AdS_d$ for
$d>4$. This problem is now under investigation. Prior it is
solved, we can only study the purely bosonic theory with totally
symmetric higher spin fields, which is the subject of this paper,
and its $\N$=$1$ supersymmetric version, which is the subject of
the forthcoming paper \cite{AV}.

\section{5d Higher Spin Algebra}
\label{5d Higher Spin Algebra}

The $AdS_5$ higher spin algebras are expected to identify with
$4d$ conformal higher spin algebras studied by Fradkin and Linetsky
\cite{FLA}, and their further extensions \cite{BHS} and
reductions \cite{SSd,BHS}.
One starts with the Lie superalgebra constructed via
supercommutators of the star product algebra (\ref{prod}). In
\cite{BHS} it was argued that this algebra as a whole,
called $hu(1,1|8)$ \cite{KV1}, may play
a key role in a $AdS_5$ higher spin gauge theory.
The set of the gauge fields corresponding to the algebra $hu(1,1|8)$ is
\be
\label{gfg}
\go(a,b|x )= \sum_{m,n=0}^\infty \frac{1}{m!n!}
\go^{\ga_1 \ldots \ga_m}{}_{\gb_1\ldots \gb_n} (x)
a_{\ga_1}\ldots a_{\ga_m}
b^{\gb_1}\ldots b^{\gb_n}\,.
\ee
The $5d$ higher spin field strength has the form
\be
\label{FS}
R (a,b|x) = d\go (a,b|x) +\go (a,b|x) *\wedge \go (a,b|x)\,.
\ee
The higher spin gauge fields in (\ref{gfg})
contain 1-forms in all representations (\ref{Xsu}). According to the
analysis of \cite{SSd} and section \ref{$su(2,2)$ - $o(4,2)$ Dictionary}
of this paper,
only the fields with $n=m$ correspond to usual (i.e., totally symmetric)
higher spin fields. Before the free theory of the mixed symmetry $AdS_5$
higher spin gauge fields is
elaborated, we confine ourselves to the higher spin algebra associated with
the simplest case of the purely bosonic theory of totally symmetric
higher spin fields.

We therefore want
to have only the gauge fields carrying equal numbers of the upper and
lower $su(2,2)$ indices. As a result, the elements of the
higher spin algebra  should  satisfy
\be
\label{en}
N_a f = N_b f\,,
\ee
where
\be
\label{Nab}
N_a = a_\gga \f{\p}{\p a_\gga }\,,\qquad N_b = b^\gga \f{\p}{\p b^\gga }\,.
\ee
This is equivalent to the condition \cite{FLA}
\be
\label{cent}
N * f = f *N\,.
\ee
Thus, the bosonic $5d$ higher spin algebra identifies
with the Lie algebra built from the star-commutators of the
elements of the centralizer of $N$ in the star product algebra
(\ref{prod}). The same algebra
(although rewritten in the $4d$ covariant notations)
was interpreted in \cite{FLA} as the $4d$ conformal higher spin algebra
called $hsc^\infty (4)$ and was proved to give rise to the
gauge invariant cubic interactions of the
$4d$ conformal higher spin theory in \cite{FL}.
We change the names of some of the
higher spin superalgebras in accordance with
the notation of \cite{KV1,BHS} to include in our systematics the
two-parametric series of matrix extensions of the higher spin
superalgebras. In particular we will use the name \hsa for the
algebra $hsc^\infty (4)$ of \cite{FLA}.

The set of the gauge field corresponding to the algebra \hsa is
\be
\label{gf}
\go(a,b|x )= \sum_{n=0}^\infty \frac{1}{(n!)^2}
\go^{\ga_1 \ldots \ga_n}{}_{\gb_1\ldots \gb_n} (x)
a_{\ga_1}\ldots a_{\ga_n}
b^{\gb_1}\ldots b^{\gb_n}\,.
\ee
The \hsa field strength has the form (\ref{FS}) and admits
analogous expansion
\be
\label{CR}
R(a,b|x )= \sum_{n=0}^\infty \frac{1}{(n!)^2} R^{\ga_1 \ldots
\ga_n}{}_{\gb_1\ldots \gb_n} (x) a_{\ga_1}\ldots a_{\ga_n}
b^{\gb_1}\ldots b^{\gb_n}\,.
\ee

So far we considered complex fields.
To impose the reality conditions let us define the
involution $\dagger$ by the relations
\be
\label{inv}
({a}_\ga )^\dagger  = i b^\gb C_{\gb \ga}\,,\qquad
({b}^\ga )^\dagger = i C^{\ga\gb} a_\gb \,,
\ee
Since an involution is required to reverse an order of product
factors
\be
\label{ord}
(f*g)^\dagger = g^\dagger *f^\dagger\,
\ee
and to conjugate complex numbers
\be
\label{alin}
(\mu f)^\dagger = \bar{\mu} f^\dagger\,,\qquad \mu \in {\bf C}\,,
\ee
the definition (\ref{inv}) contains an additional factor of $i$
compared to the complex conjugation (\ref{re}).
The involution $\dagger$ leaves invariant the defining
relations (\ref{osc}) of the star product
algebra and has the involutive property $(\dagger )^2 = Id$.
By (\ref{ord}) the action of $\dagger$ extends to
an arbitrary element $f$ of the star product algebra. Since
the star product we use corresponds to the totally symmetric
(i.e. Weyl) ordering of the product factors, the result is simply
\be
\label{dag}
(f(a_\ga , b^\gb ))^\dagger =
f (i b^\gga C_{\gga \ga} ,i C^{\gb\gga} a_\gga )\,.
\ee
It is elementary to check directly with (\ref{prod})
that (\ref{dag}) defines an involution of
the star product algebra.

The reality conditions on the elements of the higher spin algebra
have to be imposed in a way consistent with the form of the
higher spin curvature. This is equivalent to singling
out a real form of the higher spin Lie algebra.
With the help of any involution $\dagger$ this is achieved
by imposing the reality conditions
\be
\label{reco}
f^\dagger = -  f\,.
\ee
This condition defines the real higher spin
algebra $hu(1,0|8)$ for four pairs of oscillators and $cu(1,0|8)$
as its subalgebra being the centralizer of $N$.
Note that the operator $N$ is self-conjugated
\be
\label{Nd}
N^\dagger = N\,.
\ee

Let us stress that the condition (\ref{reco})
extracts a real form of the Lie superalgebra built from the
star product algebra but not of the associative star product
algebra itself. The situation is very much the same as for
the Lie algebra $u(n)$ singled out from the complex Lie algebra
of $n\times n$ matrices by the condition (\ref{reco})
with $\dagger$ identified with the
hermitian conjugation. Antihermitian matrices form the Lie algebra
but not associative algebra. In fact, the relevance of the
reality conditions of the form (\ref{reco}) is closely related with
this matrix example because it demonstrates that the spin 1
(i.e., purely Yang-Mills) part of the matrix extensions of the
higher spin algebras is compact. More generally, these reality
conditions guarantee
that the higher spin symmetry admits appropriate unitary highest weight
representations. Note that in the sector of
the $AdS_5$ algebra $su(2,2)$ the reality condition (\ref{reco})
is equivalent to (\ref{re}).

The higher spin gauge fields $\go (a,b|x) $ are required to satisfy
the condition analogous to (\ref{reco})
\be
\label{rego}
\go^\dagger = -  \go\,,
\ee
that gives rise to the component form of the reality condition
by virtue of (\ref{dag}).

For any fixed $n$  the connection
$\go^{\ga_1 \ldots \ga_n}{}_{\gb_1\ldots \gb_n} (x)$
is reducible because it is not traceless.
It decomposes into the set of $n+1$ irreducible
 components
$\go^{\prime \ga_1 \ldots \ga_k}{}_{\gb_1\ldots \ga_k}$
with all $k$ in the interval $n\geq k \geq 0$
( $\go^{\prime \ga_1 \ldots \ga_{k-1} \ggg}{}_{\gb_1\ldots \gb_{k-1}\ggg}
=0$).
As a result, fields of every spin
appear in infinitely many copies in the
expansion (\ref{gf}). The origin of this infinite degeneracy
can be traced back            to the fact that the algebra
$A_4^0$ has infinitely many ideals $I_{P(N)}$ associated with
various central elements $P(N)$ being
star-polynomials of $N$, $\{x\in I_{P(N)} : x=P(N)*y,\quad y*N=N*y\}$
\cite{FLA}. On the one hand this infinite degeneracy
makes $5d$ higher spin gauge theories
 reminiscent of the superstring theory that contains infinitely
many (massive) modes of any given symmetry type. On the other hand
a question arises whether it is possible to consider
consistent higher spin
models with reduced spectra of spins associated with
the quotient higher spin algebras. The most interesting reductions
are provided with the algebra $hu_0(1,0|8)$=\hsa$\!/I_N$ called
$hsc^0 (4)$ in \cite{FLA} and its further reduction $ho_0(1,0|8)$
\cite{BHS} called $hs(2,2)$ in \cite{SSd}. ($I_N$ is the ideal spanned
by the elements of the form $g=N*f = f*N$.) The gauge fields of the algebra
$hu_0(1,0|8)$ correspond to the set of all integer spins $s\geq 1$
(every spin appears once) while $ho_0(1,0|8)$ describes its reduction
to the subalgebra associated with even spins.
As we show both options are allowed in the framework of the
cubic analysis  of this paper. We start in the
section \ref{d5 Higher Spin Gauge Fields} with the analysis of the
unreduced case of \hsa considering the reduced cases
afterwards in section \ref{Maximally Reduced Model}.
Note that from this perspective our conclusions
are somewhat different from those of \cite{FL} where it was argued that
only the unreduced algebra \hsa admits consistent dynamics in the
framework of the $4d$ conformal higher spin theory.
{}From the perspective of AdS/CFT correspondence the most
interesting cases are associated either with the maximally reduced
models \cite{SSd,BHS} and their supersymmetric extensions or
with the unreduced models based on the algebras $hu(m,n|8)$
\cite{BHS} which, presumably,  give rise to all types of
$AdS_5$  massless fields.

\section{5d Higher Spin Gauge Fields}
\label{d5 Higher Spin Gauge Fields}

The \hsa linearized higher spin curvature
\be
\label{lin}
R_1 (a,b|x) = d \go_1(a,b|x)  +\go_0 (a,b|x) * \wedge \go_1 (a,b|x)+
\go_1 (a,b|x) * \wedge \go_0 (a,b|x)\,,
\ee
with
\be
\label{go0}
\go_0 (a,b|x) =  \go_0{}^\ga{}_\gb a_\ga b^\gb\,,\qquad \go_0{}^\ga{}_\ga =0\,
\ee
satisfying the zero curvature condition (\ref{ads5}),
provides the $5d$ spinor version of the formula (\ref{R1A}).
Equivalently,
\be
\label{llin}
R_1 (a,b|x) = d \go_1 (a,b|x)  +\go_0{}^\ga{}_\gb (x)
(\f{\p}{\p b^\ga}b^\gb -a_\ga \f{\p}{\p a_\gb})
\wedge \go_1 (a,b|x)\,.
\ee
The component formula reads
\bee
R_1{}^{\ga_1 \ldots \ga_{s-1}}{}_{\gb_1 \ldots \gb_{s-1}} &=&
d\go{}^{\ga_1 \ldots \ga_{s-1}}{}_{\gb_1 \ldots \gb_{s-1}}
- (s-1)\Big ( \go_0{}^{\{\ga_1}{}_\gga \wedge
\go{}^{\gga \ga_2 \ldots \ga_{s-1}\}}{}_{\gb_1 \ldots \gb_{s-1}}\nn\\
&-&\go_0{}^{\gga}{}_{\{\gb_1}\wedge
\go{}^{ \ga_1 \ldots \ga_{s-1}}{}_{\gga \gb_2 \ldots \gb_{s-1}\}}
\Big )\,.
\eee

The linearized (Abelian) higher spin gauge transformations are
\bee
\label{lintr1}
\delta \go_0 (a,b|x) = D_0 \gvep(a,b|x)\,,
\eee
where
\bee
\label{CD0}
D_0 = d   +\go_0{}^\ga{}_\gb
(\f{\p}{\p b^\ga}b^\gb -a_\ga \f{\p}{\p a_\gb})\,
\eee
is the background  covariant derivative.
The fact that $\go_0$ satisfies the zero-curvature condition
implies
\be
\label{Rinv}
\delta_0 R_1 =0 \,.
\ee

To decompose
the representations of the $AdS_5$ algebra $su(2,2)\sim o(4,2)$
into representations of its Lorentz subalgebra $o(4,1)$
we use the antisymmetric compensator $V^{\ga\gb}$.
A $su(2,2)$ counterpart of the reduction of the
tensor higher spin gauge field
$dx^\un \go_\un{}^{A_1 \ldots A_{s-1}, B_1\ldots B_{s-1} }$
carrying the irreducible representation of the $AdS_d$ algebra $o(d-1,2)$
described by the traceless two-row Young diagram of length $s-1$ into
a collection of the Lorentz covariant higher spin  1-forms
$dx^\un \go_\un{}^{a_1 \ldots a_{s-1}, b_1\ldots b_t }$
with all $0\leq t \leq s-1$  goes as follows.
The field $V^{\ga\gb}$ is used to raise and lower spinor indices. Then, the
Lorentz algebra
irreducible components correspond to various
types of symmetrization between the two types of indices,
i.e. again to all two-row traceless
Young diagrams but now in the spinor indices,
$\go^\prime_{\ga_1 \dots \ga_{s-1+q}, \gb_1\ldots \gb_{s-1-q}}$ with all
$0\leq q\leq s-1$ (all traces with $V^{\ga\gb}$ are zero and
symmetrization with respect to any $s+q$ indices gives zero).
The identification with the $o(4,1)$ tensor notation is
\be
\label{sviden}
\go^\prime_{\ga_1 \dots \ga_{s-1+t}, \gb_1\ldots \gb_{s-1-t}} \sim
\go^{a_1 \ldots a_{s-1}, b_1\ldots b_{t} }\,.
\ee
For example,
the two-row rectangular diagram of length $s-1$ in tensor notation is
described by the one-row diagram of length $2(s-1)$ in the spinor
notation, while the two-row rectangular diagram of length $s-1$ in
spinor notation corresponds to the one-row diagram of length $s-1$
in the tensor notation. (Particular manifestations of this relationship
are those between the vector and traceless antisymmetric second-rank
spinor or antisymmetric tensor and  symmetric second rank spinor,
both underlying the isomorphism between the spinor and vector
realization of the $5d$ space-time symmetry algebras.)  Note that
the analogous identification of the representations was discussed in
the recent paper \cite{SSd}, where the spinor version of the
linearized higher spin curvatures has been presented. The difference is
that in this paper we use the manifestly $su(2,2)$ covariant
compensator formalism that simplifies
greatly the analysis of the interactions.

In what follows we shall use the two sets of the
differential operators in the spinor variables
\be
\label{S}
S^- =V^{\ga\gb} a_\ga \f{\p}{\p b^\gb} \,,\qquad
S^+ =V_{\ga\gb} b^\ga \f{\p}{\p a_\gb} \,,\qquad
S^0 = N_b - N_a
\ee
and
\be
\label{T}
T^+ = a_\ga b^\ga\,,\qquad T^- =\frac{1}{4}
 \frac{\partial^2 }{ \partial a_\ga
\partial b^\ga } \,,\qquad T^0 =\frac{1}{4}
( N_a +N_b  +4 )\,.
\ee
They form two mutually commuting $sl_2$ algebras
\be
\label{sl2S} [ S^0 , S^\pm ] =  \pm  2 S^\pm \,,\qquad
[S^- , S^+ ] =  S^0\,,
\ee
\be
\label{sl2inv}
[ T^0 , T^\pm ] =  \pm  \half T^\pm \,,\qquad
[T^- , T^+ ] =  T^0\,,
\ee
\be
\label{TS}
[ T^i , S^j ] = 0\,.
\ee
(Unusual normalization of the generators $T^i$ in (\ref{T}),
and (\ref{sl2inv}) is chosen for the future convenience).

The operators $T^i$ and $S^0$ are  independent of
the compensator $V^{\ga\gb}$ and, therefore, are $su(2,2)$ invariant.
As a result,
\be
\label{DT0}
D_0 (T^i ) =0\,,\qquad D_0 (S^0 ) =0\,.
\ee
(These relations have to be understood in the sense
that $D_0 (X(f)) = X(D_0 (f))$, where $X$ is one of the operators
$T^i$ and $S^0$, while $f$ is an arbitrary element of the
 star product algebra.)
A useful consequence of this fact is
\be
\label{RT}
R_1 (T^j (\go )) = T^j (R_1 ( \go ) )\,.
\ee
{}From (\ref{sl2S}) it also follows
\be
\label{SDS}
[S^+ , D_0 (S^- )]+ [D_0 (S^+ ), S^- ] =0\,.
\ee
According to (\ref{en}) the elements of the higher spin algebra
\hsa satisfy $S^0 (f) =0$, i.e.
\be
\label{S0go}
S^0 \go (a,b |x )=0 \,,\qquad S^0 R(a,b |x )=0\,.
\ee
As a result, the operators $S^+$ and $S^-$ commute to each other  on the
higher spin gauge fields and field strengths of \hsa.

The $V^{\ga\gb}$ - dependent operators $S^\pm$ are only Lorentz invariant.
In accordance with (\ref{h}) and (\ref{tr1})
\be
D_0 S^- = h^{\ga\gb} a_\ga \f{\p}{\p b^\gb }\,,\qquad
D_0 S^+ = -h_{\ga\gb} b^\ga \f{\p}{\p a_\gb }\,.
\ee
Let us note that the background covariant derivative $D_0$ (\ref{CD0})
admits the representation
\be
\label{D0a}
D_0 = D_0^L + \half [ S^- , D_0 S^+ ]
= D_0^L -\half h^\ga {}_\gb (a_\ga\f{\p}{\p a_\gb} - b^\gb\f{\p}{\p b^\ga})\,,
\ee
where the background Lorentz derivative $D_0^L$ commutes with all
operators $T^i$ and $S^i$.

{}From the star product (\ref{prod}) it follows that
\be
\label{N*}
N * f =  \left (T^+ -T^-  +\half ( N_b - N_a)\right ) f \,.
\ee
According to (\ref{en}), for $f \in$\hsa  this simplifies to
\be
\label{NT}
N * f = f* N = (T^+ - T^- ) f\,.
\ee

The decomposition into $su(2,2)$ irreducible fields is
\be
\label{irf}
\go(a ,b ) = \sum_{s,n=0}^\infty (T^+ )^n v_n (T^0)
\go^s_n (a,b)
\,,
\ee
with
\be
\label{T0s}
T^0\, \go^s_n = \frac{1}{2}(s+1) \, \go^s_n\,,
\ee
\be
\label{hw}
T^- \go^s_n (a,b) =0\,.
\ee
For the future convenience, we fix the normalization coefficients
$v_n (T^0 )$ in (\ref{irf}) in the form
\be
\label{coe} v_n (\half (s+1))  = v(\half (s+1)) (2i)^n
\sqrt{\f{(2s+1) !} {n! (n+ 2s+1)!}}\,.
\ee
Note that the factor of $i^n$ in (\ref{coe}) implies that the different
copies of the fields $\go_n^s$ with the same spin contained in polynomials of
degree $4p$ and $4p+2$ contribute with opposite signs. This is appropriate
because the coefficients in front of the corresponding parts
of the invariant action will be shown to have opposite signs as well.

Due to (\ref{RT}), the linearized curvatures admit the expansion
analogous to (\ref{irf})
\be
\label{ircu}
R_1 (a ,b ) =  \sum_n  ( T^+ )^n   v_n ( T^0 )
          R_{1,n} (a ,b )
\ee
with
\be
\label{T-R}
T^- R_{1,n} (a ,b ) =0\,.
\ee

Let us now explain how the invariant version of the Lorentz
covariant decomposition used in \cite{LV,SSd} can be defined.
Lorentz multispinors associated with the two-row Young
diagrams having $n_1$ and $n_2$ cells in the upper and lower rows respectively
($n_1 \geq n_2$) can be described as the polynomials $\eta (a, b)$ of the
spinor variables  $a_{\ga}$ and $b^{\gb}$ subject to the conditions
\be
N_a \eta (a, b )=n_1\eta (a, b )\,,\qquad N_b \eta (a, b )=n_2\eta (a, b
)\,,\qquad
\ee
\be
\label{YD}
S^- \eta (a, b )= 0\,,
\ee
where the latter condition implies that the symmetrization over any $n_1 +1$
indices gives zero. The tracelessness condition reads in these terms
\be T^- \eta (a, b )=0 \,.
\ee

The $su(2,2)$ irreducible higher spin gauge field $\go$
admits the following representation in terms of the
Lorentz - irreducible higher spin fields
\be
\label{lgf}
\go (a , b ) = \sum_{t=0} (S^+ )^t \eta^t (a , b)\,,
\ee
(Note that the asymmetric
form  of this  formula with respect to $a_\ga$ and $b^\gb$
is a result of a particular basis choice.) Since
$\go (a , b )$ has equal numbers of $a$ and $b$, we set
$ 2t = n_1 - n_2$. For the spin $s$  we have
$2(s-1) =n_1 + n_2$ (cf. (\ref{T0s})).
For  $s$ fixed,  $t$ ranges from $0$ to $s-1$.

One can treat the Lorentz-irreducible 1-forms $\eta_t (a ,b )$ as
an alternative  basis of the higher spin gauge
fields. The linearized  higher spin curvature 2-forms (\ref{lin})
admit the analogous expansion
\be
R_1 (a , b ) = \sum_{t=0} (S^+ )^t
r_{1}^t (a , b)\,,
\ee
with the Lorentz - irreducible component curvatures
$r_{1}^t (a , b)$ satisfying the Young property
\be
S^-  r_{1}^t (a, b )= 0\,
\ee
and the tracelessness condition
\be
T^- r_{1}^t (a, b )=0 \,.
\ee

{}From the definition of $r_{1}^t (a , b)$ it follows  that
\be
\label{r1t}
r_{1}^t (a , b) = D_0^L \eta^t (a,b) + \tau_- (\eta^{t+1} (a,b))
+\tau_+ (\eta^{t-1} (a,b) )\,,
\ee
where $D^L_0$ is the Lorentz covariant derivative and
the $5d$ spinor realization of the  operators (\ref{t-})
and (\ref{t+}) is
\be
\label{ts+}
\tau_+ = \f{1}{S^0 +1} D_0 (S^- )\,,
\ee
\bee
\label{ts-}
\tau_-  =  \half \Big ( S^+ [S^- , D_0 (S^+ )] -D_0 (S^+ )S^0
-(S^+ )^2 \frac{1}{S^0 +1} D_0 ( S^- ) \Big )\,.
\eee
One can see that the properties (\ref{tau+-}) are satisfied
on the space of functions $\eta$ satisfying (\ref{YD}).

Analogous decomposition
\be
\label{DDD}
D_0 = D_0^L + \tau_+ +\tau_-
\ee
exists in the original basis of fields $\go$ satisfying the condition
(\ref{S0go}).
The explicit form of $\tau_\pm$ in this basis is
\be
\label{ct}
\tau_\pm = \f{1}{4} \left
( [S^- , D_0  S^+ ]  \pm \f{1}{\sqrt{1-4S^+ S^- }} \Big ([S^- , D_0  S^+ ]
+2 D_0 (S^+ S^- )\Big )\right)\,.
\ee
Derivation of this formula is  more complicated.
It is based on the fact that the operator $S^- S^+ = S^+ S^-$
diagonalizes on the vectors with different $t$ in
(\ref{lgf}) with the eigenvalues $-t(t+1)$ so that the operator $\hat{t}$
\be
\hat{t} = \half (\sqrt{1-4S^+ S^- } -1 )
\ee
has eigenvalues $t$. The property that
\be
 [\hat{t} , \tau_\pm ] = \pm \tau_\pm
\ee
turns out to be equivalent to
\be
\label{S+-t}
[S^+ S^- , \tau_\pm ] = (1 \mp \sqrt{1-4S^+ S^- }) \tau_\pm\,.
\ee
Taking into account the fact that the decomposition of $D_0$ into
eigenspaces $D_0^L$, $\tau_+$ and $\tau_-$ of $S^+ S^-$ is unique,
the problem is to find such operators $\tau^\pm$
on the space of functions satisfying (\ref{S0go})
that the formulas (\ref{S+-t}) and (\ref{DDD}) are true.
Formula (\ref{ct}) solves this problem. For
(\ref{DDD}) this is obvious.
The verification of the formula (\ref{S+-t}) is also elementary
with the help of identities valid on the subspace of null-vectors of
$S^0$
\be
[S^+ S^- , [S^- , D_0 S^+ ] ] =-2D_0 (S^+ S^- )\,,
\ee
\be
  [S^+ S^- , D_0 (S^+ S^- ) ] = 2\left ( D_0 (S^+ S^- )
+S^+ S^- [S^- , D_0 S^+ ] \right )\,.
\ee
Another useful fact is that the operator
\be
\tau_0 = S^- D_0 (S^+ ) - S^+ D_0 (S^- )
\ee
does not affect the gradation $t$, i.e.
\be
[S^+ S^- , \tau_0 ] =0\,.
\ee
It is less trivial to check that $(\tau^\pm )^2 =0$. The simplest
way is  to use the basis of the fields $\eta_t$, i.e. the operators
$\tau_\pm$ in the form (\ref{ts+}) and (\ref{ts-}).

Let us note that the variables $\go$ and $\eta$ can be interpreted
as different representatives of the same representation  of the
$sl_2$ algebra spanned by the operators $S^j$. Namely, the variables
$\go$ are associated by (\ref{T0s})
with the elements having zero eigenvalue of the
Cartan element, while the variables $\eta$ are associated
by (\ref{YD}) with the lowest
weight vectors. This suggests the idea that there should be some
formulation operating in terms of the representations of this $sl_2$
algebra as a whole.

Equipped with the operators $\tau_\pm$ and $\tau_0$,
one can write the spinor form of the constraints
(\ref{conv}) either as
\be
\label{const0}
\tau_0 \wedge \tau_+ R_1 =0\,
\ee
in the basis $\go$ or as
\be
\label{const-}
\tau_0 \wedge \tau_+ r_1 =0\,
\ee
in the basis $\eta$. To obtain the spinor form of the First  On-Mass-Shell
theorem one takes into account that, as shown in the beginning of this
section  (see also \cite{SSd}), the Weyl tensor,
described in terms of tensors by the   length $s$
two-row traceless Young diagram
            $C^{a_1 \ldots a_s}{}_,{}^{b_1 \ldots b_s}$,
is described in terms of
spinors by a rank $2s$ totally symmetric multispinor
$C^{\ga_1 \ldots \ga_{2s}}$. Since the First
On-Mass-Shell Theorem (\ref{CMT}) is true for any irreducible
higher spin field in the expansion (\ref{ircu}), it acquires the
form
\be
\label{scomt}
R_{1,n} (a ,b )\Big |_{m.s.} =
H_{2\ga}{}^\gb \f{\p^2}{\p a_\ga \p b^\gb }
Res_\mu (C_n (\mu a + \mu^{-1} b) )\,,
\ee
where the label $\Big |_{m.s.}$
implies the on-mass-shell consideration modulo terms proportional to
the left hand sides of the free field equations and constraints
(\ref{const0}) (equivalently, (\ref{const-})).
$Res_\mu $ singles out the $\mu-$independent part of
a Laurent series in $\mu$, i.e.
\be
\label{res}
Res_\mu
\Big (\sum_{n=-\infty}^\infty \ga_n \mu^n \Big ) =
\f{1}{2\pi i}\oint d \log \mu
\Big (\sum_{n=-\infty}^\infty \ga_n \mu^n \Big )
= \ga_0\,.
\ee
Note that a function of one spinor variable
\be
C_n (\mu a + \mu^{-1} b) =
\sum_{k,l} \f{\mu^{k-l}}{k! l!}
C_n^{\ga_1 \ldots \ga_{k}\gb_1 \ldots \ga_{l}}
a_{\ga_1} \ldots a_{\ga_k} b_{\gb_1} \ldots b_{\gb_l}
\ee
has totally symmetric coefficients
$C^{\ga_1 \ldots \ga_{k}\gb_1 \ldots \gb_{l}}$ while $Res_\mu$
singles out its part containing equal numbers of the oscillators
$a$ and $b$ that belongs to \hsa.

\section{Central On-Mass-Shell Theorem}
\label{5d Central On-Mass-Shell Theorem}

The matter fields and higher spin Weyl tensor can be interpreted as
representatives of the $\sigma_-$ cohomology group associated with
the so-called twisted adjoint representation of the higher spin algebra.
Given automorphism $\tau$ of the higher spin algebra (in fact any
associative algebra used to build a Lie superalgebra via supercommutators),
one defines the covariant derivative $\tilde{D}$ of a field $C$
taking values in the twisted adjoint representation
\be
\D C = d C +\go * C - C*\tau (\go )\,.
\ee
The property that $\tau$ is an automorphism guarantees that this definition
is consistent with the Bianchi identities. (See \cite{Ann,gol} for particular
examples and references.) To have a formulation
in terms of Lorentz covariant fields (i.e. finite-dimensional
representations of the Lorentz algebra),
$\tau$ is required to leave invariant the Lorentz subalgebra of the
full $AdS$ algebra. In terms of the compensator formalism this is
automatically achieved by using the compensator field for definition of
$\tau$. For the problem under consideration, the appropriate definition
is
\be
\label{tau5}
\tau (a_\ga ) = b^\gb V_{\gb \ga} \,,\qquad
\tau (b^\ga ) =       V^{\ga \gb} a_\gb\,
\ee
implying
\be
\tau (f(a,b|x)) = f(\tau (a),\tau (b) |x)\,.
\ee

Let us note that in this section we require the compensator
$V_{\ga\gb}$ to be a constant so that $\tau$ commutes with
the exterior differential $d$.

The linearized covariant derivative (\ref{D0a})
in the adjoint representation can be written as
\be
D_0 = D_0^L +\half [h^\ga{}_\gb a_\ga b^\gb ,\,.\, ]_* \,.
\ee
Analogously to the $4d$ case \cite{Ann,gol}, the twisted
linearized covariant derivative results from the replacement
of the star commutator to star anticommutator
in the part of the covariant derivative associated with the frame 1-form
\be
\label{cD0}
\D_0 (C) = D_0^L (C) +\half \{ h^\ga{}_\gb a_\ga b^\gb  , C\}_* \,.
\ee
In fact, this is not surprising because
the only nontrivial Lorentz covariant definition
of the restriction of $\tau$ to the $AdS_d$ algebra in any dimension
is to change a sign of the $AdS$ translations. {}From the perspective
of the higher spin symmetry the problem therefore is to find an appropriate
extension of this automorphism of the $AdS$ algebra to the full higher
spin algebra. This is achieved by the definition (\ref{tau5}) for the
case of $AdS_5$. For some specific choice of the compensator,
this definition reproduces the twisted adjoint
representation used in \cite{SSd}.

The twisted covariant derivative (\ref{cD0}) has the  form
\be
\D_0 (C) = D_0^L (C) +\sigma_- +\sigma_+\,,
\ee
where
\be
\label{sig}
\sigma_- =- \f{1}{4} h^\ga{}_\gb   \f{\p^2}{\p a_\gb \p b^\ga} \,,\qquad
\sigma_+ = h^\ga{}_\gb  a_\ga b^\gb\,.
\ee
The operators $D^L_0$ and $\sigma_\pm$ have the properties
\be
\label{ds}
(\sigma_\pm )^2 =0\,,\qquad (D_0^L )^2 + \{\sigma_- , \sigma_+ \} =0\,,
\qquad \{ D_0^L , \sigma_\pm \} =0\,.
\ee
Only the operator $D_0^L$ acts nontrivially
(differentiates) on the space-time coordinates while $\sigma_\pm$
act in the fiber linear space $V$ isomorphic as a linear space
to the twisted adjoint representation of the higher spin algebra.
Also there is
the gradation operator $G=\half (N_a +N_b )$ such that
\be
[G , D_0^L ] = 0\,,\qquad [G ,\sigma_\pm ] = \pm \sigma_\pm\,.
\ee
Since $V$ is spanned by polynomials in the spinor variables
$a_\ga$ and $b^\gb$,
the spectrum of $G$ in $V$ is bounded from below.

The important observation is (see, e.g., \cite{SVsc}) that
the nontrivial dynamical equations hidden in
\be
\label{geneq}
\tilde{D}_0 (C) =0
\ee
are in the one-to-one correspondence with the nontrivial
cohomology classes of $\sigma_-$. For the case
with $C$ being a
0-form, the relevant cohomology group is $H^1 (\sigma_- )$.
For the more general situation with $C$ being a $p$-form,
the relevant cohomology group is $H^{p+1} (\sigma_- )$.
From this perspective, the operator $\tau_-$ identifies with
$\sigma_-$ in the sector of the higher spin gauge 1-forms.

Indeed, consider the decomposition of the space of fields $C$
into the direct sum of eigenspaces of $G$. Let a
field having a definite eigenvalue $k$ of $G$ be
denoted $C_k$, $k= 0,1,2 \ldots$. Suppose that the dynamical
content of the equations (\ref{geneq}) with the eigenvalues
$k \leq k_q $ is found. Applying the operator $D_0^L +\sigma_+$
to the left hand side
of  the equations (\ref{geneq}) at  $k \leq k_q $ we obtain
taking into account (\ref{ds})
that
\be
\label{sdc}
\sigma_- (D_0^L +\sigma_- +\sigma_+ )( C_{k_q+1} ) =0\,.
\ee
Therefore $(D_0^L +\sigma_- +\sigma_+ )( C_{k_q +1} )$ is
$\sigma_-$ closed. If the group $H^1 (\sigma_- )$ is trivial
in the grade $k_q+1$
sector,  any solution of (\ref{sdc}) can be written in
the form
$(D_0^L +\sigma_- +\sigma_+ )( C_{k_q+1} ) = \sigma_- \tilde{C}_{k_q +2}$
for some  field $\tilde{C}_{k_q +2}$. This,
in turn, is equivalent to the statement that one can adjust
$C_{k_q +2}$ in such a way
that $\tilde{C}_{k_q +2} =0$ or, equivalently, that the part of the
equation (\ref{geneq}) of the grade $k_q+1$ is some constraint
that expresses $C_{k_q +2}$ in terms of the derivatives of $C_{k_q +1}$
(to say that this is a constraint we have used the assumption that
the operator $\sigma_-$ is algebraic
in the space-time sense, i.e. it does not contain space-time derivatives.)
If $H^1 (\sigma_- )$ is nontrivial, this means that the equation
(\ref{geneq}) sends the corresponding cohomology class to zero and,
therefore, not only
expresses the field $C_{k_q+2}$ in terms of  derivatives of
$C_{k_q+1}$ but also imposes some additional
differential conditions on $C_{k_q+1}$.  Thus, the nontrivial
space-time differential equations described by (\ref{geneq})
are classified by the cohomology group $H^1 (\sigma_- )$.

The nontrivial dynamical fields are associated with
$H^0 (\sigma_-)$ which is always non-zero because it at least
contains a nontrivial subspace of $V$ of minimal
grade. As follows from the $H^1 (\sigma_- )$
analysis of the dynamical equations, all
fields in $V/H^0 (\sigma_-)$ are auxiliary, i.e. express via the
space-time derivatives of the dynamical fields by virtue
of the equations (\ref{geneq}).

In the problem under consideration we are interested in the
sector of fields $C(a,b |x)$ that commute to $N$
(i.e. $N_a (C) = N_b (C)$). In this sector the representatives of
$H^0 (\sigma_- )$ (i.e., fields $C$ satisfying $\sigma_- (C)=0$)
are described by the fields of the form
\be
C_{0} (a,b |x) =  Res_\mu C_0 (\mu a +\mu^{-1} b, a_\ga b^\ga|x)\,.
\ee
We see that these are just the fields that appeared in the first
on-mass-shell theorem (\ref{scomt}). The additional dependence on
$a_\ga b^\ga$ matches the degeneracy of the higher spin
fields of \hsa due to traces (i.e., ideals generated by $N$).

Application of the same analysis to the higher spin gauge 1-forms
with the operator $\tau_-$ instead of $\sigma_-$ leads to the
following interpretation of the results of section
\ref{d5 Higher Spin Gauge Fields}. The dynamical fields with spins
$s\geq 1$ belong to the cohomolgy group $H^1 (\tau_- )$.
$\tau_- $ exact 1-forms $\go (a,b|x)= \tau_- (\xi )$ describe pure gauge
degrees of freedom in $\go (a,b|x)$ analogous to the antisymmetric part of the
frame field associated with the local Lorentz transformations in gravity. The
cohomology group $H^2 (\tau_- )$ responsible for nontrivial differential
conditions on the higher spin gauge fields is a direct sum of two linear spaces
\be
H^2 (\tau_- ) = V_2^{E} (\tau_- ) \oplus V_2^W (\tau_- )\,.
\ee
The space $V_2^W (\tau_- ) $ called Weyl cohomology is spanned by the
2-forms of the form of the right hand side of the equation (\ref{scomt}),
i.e. a generic element of $V_2^W (\tau_- ) $ has the form
(to simplify formulae, in the rest of this section we confine ourselves to
the case of irreducible fields of different spins satisfying $T^- \go =0$)
\be
\label{H2W} H_{2\ga}{}^\gb \f{\p^2}{\p a_\ga \p b^\gb }
Res_\mu (C (\mu a + \mu^{-1} b) )\,.
\ee
The space $V_2^E (\tau_- ) $ called Einstein cohomology is spanned by the
2-forms of the form
\be
H_{2\ga}{}^{\gb} \Big (\f{\p^2 }{\p a_\ga \p b^\gb } R(a,b) +
b^\ga a_\gb r(a,b ) \Big )\,,
\ee
where the 0-forms $R(a,b)$ and $r(a,b)$ have themselves the
properties of the dynamical fields, i.e.
\be
S^\pm R=0\,,\q T^- R =0\,,\q
S^\pm r=0\,,\q T^- r =0\,.
\ee
The 0-forms $R$ and $r$ parametrize the right hand sides of the
spin $s\geq 2$ equations of motion. They generalize the traceless part
of the Ricci tensor and the scalar curvature, respectively. In other words,
they correspond to the dynamical equations of motion associated with
the irreducible traceless parts of the double traceless Fronsdal fields
$\varphi_{a_1\ldots a_n}$ in the action (\ref{fract}). The equation
(\ref{scomt}) sends the right hand sides of the dynamical equations
associated with the $R$ and $r$ to zero imposing no other conditions
on the dynamical fields because the Weyl cohomology remains arbitrary.
This is the content of the first of-mass-shell theorem that states that
(\ref{scomt}) is equivalent to the free equations of motion for all spins
$s\geq 2$. Note that the spin 1  Maxwell  equations are not contained in
the equation (\ref{scomt}) which merely defines the associated spin 1
field strength as the degree two part of the Weyl cohomology $C(a,b)$.
The degree zero part $C(0,0)$ associated with spin 0 field does not
show up in the Weyl cohomology because the scalar field $C(0,0)$
is not associated with the gauge fields.

The fact that the equation (\ref{scomt}) sew the 0-forms $C(a,b|x)$
to the higher spin curvatures has two effects. First, what looked like
an independent dynamical spin $s\geq 1$ field in the module
$C(a,b|x)$ becomes an auxiliary field expressed by (\ref{scomt})
in terms of the dynamical fields described by the 1-form gauge fields.
Second, the fields on the right hand side of (\ref{scomt})
have to satisfy
some differential restrictions as a consequence of the Bianchi identities.
For all spins $s\geq 2$ these differential restrictions are equivalent to
what looked like independent equations in the condition that the
section $C(a,b|x)$ is flat. In other words, the Bianchi identities send
to zero the part of the cohomology group $H^1 (\sigma_- )$ associated with
all spins $s\geq 2$ ($s\geq 3/2$ when fermions are included \cite{AV}).
For spin 1 only a half of the corresponding part of $H^1 (\sigma_- )$
is sent to zero by the Bianchi conditions. This is associated with
the Maxwell equation that encodes the Bianchi identities for the
field strength expressed in terms of the 1-form potential. The
dynamical part of the Maxwell equations
is imposed by the covariant constancy condition for the
spin 1 part of $C(a,b|x)$, i.e. by setting to zero the rest of the
restriction of $H^1 (\sigma_- )$ to the spin 1 sector. The equation for
spin 0 is the condition that $H^1 (\sigma_- )=0$ in the spin 0
sector \cite{SVsc} (the situation with spin 1/2 is analogous \cite{AV}).

As a result,
we arrive at the Central On-Mass-Shell Theorem that states that the
equations (\ref{scomt}), (\ref{geneq}) describe the equations of motion
for free massless fields of all spins
along with an infinite set of constraints that express some auxiliary
fields via higher derivatives of the dynamical fields associated with
the cohomology group $H^1 (\tau_- )$ and the scalar field
$c(x) = C(a,b|x)$. Let us note that, by construction,
 the set of fields $\go (a,b |x)$ and $C(a,b|x)$ provide the complete
basis for all combinations of derivatives of massless fields of all spins
that are allowed to be nonzero by field equations (equivalently, to take
arbitrary values at any fixed point $x_0$ of space-time).
The Central On-Mas-Shell Theorem is the starting point for
the description of the nonlinear higher spin dynamics in the unfolded form.
The equation (\ref{scomt}) also plays the key role in the analysis of
cubic higher spin interactions at the action level.

The proof and the meaning of the  tensor form of the
Central On-Mass-Shell Theorem
(\ref{cmt1}) and (\ref{cmt2}) in any dimension is analogous.

\section{5d Higher Spin Action}
\label{Higher Spin Action}

The aim of this section is to formulate the action for the
totally symmetric
gauge massless boson fields in $AdS_5$ that solves the problem of
higher-spin-gravitational interactions in the first nontrivial
order. The results reported here  extend
the $4d$ results of \cite{Fort1,FV1} to $d=5$.

We shall look for the action of the form
\be
S=S_2 +S_3 +\ldots
\ee
within the perturbation expansion  (\ref{go01}) with
the background gravitational field being of the zero
order and the higher spin fields of the first order.
$S_2$ is the quadratic action that describes properly
the free higher spin dynamics. $S_3$ is the cubic part.
Higher-order corrections do not contribute to the order under
investigation.
The gauge transformations are supposed to be of the form
(\ref{godeftr}). Equivalently one can expand
\be
\delta \go = \delta_0 \go + \delta_1 \go+\ldots \,,
\ee
where
$\delta_0 \go $ is the linearized Abelian
transformation (\ref{lintr})
while $\delta_1 \go $ contains terms linear in the dynamical
fields $\go_1$. Recall that the background field $\go_0$ is chosen
in such a way that $R_0=0$ (thus implying $AdS_5$ background) so that
$R$ starts with the first order part. As a result, the deformation terms
$\Delta (R, \gvep )$ in (\ref{godeftr}) contribute to $\delta_1 \go $.

The free higher spin action $S_2$ is required to be invariant under the
linearized higher spin gauge transformations
\be
\delta_0 S_2 =0 \,.
\ee
This means that the part of the variation of the action, which is
linear in the dynamical fields, is zero. The first nontrivial
part is therefore bilinear in the dynamical fields
\be
\label{gv}
\delta_1 S = \delta_0 S_3 +\delta_1 S_2 \sim \go_1^2 \gvep\,.
\ee
Our aim is to find an action $S_3$ that admits a nontrivial
deformation of the gauge transformation guaranteeing that the gauge
variation (\ref{gv}) is zero.

Using the decomposition (\ref{godeftr}) for the gauge variation
one can rewrite the condition $\delta_1 S=0$ in the equivalent form
\be
\label{invc}
0= \delta^g S +\Delta S_2 +O( \go_1^3 \gvep )\,,
\ee
where $\delta^g$ is the original higher spin gauge transformation
(\ref{gotr}) that contains the zero-order part of the variation
along with some part of the first-order terms. Other possible
linear terms in the variation are contained in $\Delta \go_1$.
Since
\be
\Delta S_2 = \frac{\delta S_2}{\delta \go_{dyn}} \Delta \go_{dyn}
\ee
a (local) deformation $\Delta \go_{dyn}$ fulfilling
the invariance condition (\ref{invc}) exists iff
\be
\label{Y}
\delta^g S =- Y (\go_1, \frac{\delta S_2}{\delta \go_{dyn}},\gvep )
+O( \go_1^3 \gvep )\,,
\ee
where $Y (\go_1, \frac{\delta S_2}{\delta \go_{dyn}},\gvep )$ is some
trilinear local functional, i.e. iff the original gauge variation of
$S_2 + S_3$ vanishes on-mass-shell
$\frac{\delta S_2}{\delta \go_{dyn}}=0$.

Note that a deformation of the gauge variation of the extra and
auxiliary fields  does not contribute into the variation to the order under
consideration because the variation of $S_2$ with respect to these fields is
either  identically zero by the {\it extra field decoupling condition}
(\ref{extvar}) for extra fields or zero by virtue of constraints (i.e., by the
1.5-order formalism) for the Lorentz-type auxiliary fields.
This is important because the constraints for the extra and auxiliary fields
are not invariant under the original higher spin gauge transformations
$\delta^g \go$. As a result, the higher spin gauge transformation
for the extra and auxiliary fields should necessarily be deformed
to be compatible with the constraints. This phenomenon
does not however affect our consideration
because the constraints are formulated
in terms of the higher spin curvatures and therefore are invariant under
the linearized higher spin gauge transformations in the lowest order.
As a result, the deformation of the transformation low for the extra and
auxiliary fields due to the
constraints is at least of order $\go_1 \gvep$ which was argued
to be irrelevant in the approximation under consideration.

Our analysis of the gauge invariance will be based heavily on
the First On-Mass-Shell Theorem (\ref{CMT}) in its spinor form
(\ref{scomt}). Namely, the
variation $\delta^g S $ is some bilinear functional of the higher spin
curvatures $R$ which can be replaced by the linearized curvatures
$R_1$ at the order of interest. Assuming that the constraints
for auxiliary and extra fields are satisfied we can use the representation
(\ref{CMT}) for the linearized curvatures. All terms contained in
$X$ are proportional to the left hand sides of the free field equations
and, therefore, give rise to some variation of the form (\ref{Y})
that can be compensated by an appropriate deformation $\Delta \go_1$
(that itself is at least linear in the higher spin curvatures).
The terms that cannot be compensated this way are those bilinear
in the higher spin Weyl tensors $C_{A_1 \ldots A_{s-1},B_1 \ldots B_{s-1}}$.
Therefore, the condition that the higher spin action
is invariant under some deformation of the higher spin gauge
transformations is equivalent to the condition that the original
(i.e. undeformed) higher spin gauge variation of the action is zero
once the linearized higher spin curvatures $R_1$ are replaced by
the Weyl tensors $C$ according to (\ref{CMT}) i.e., schematically,
\be
\label{ncon}
\delta^g S \Big |_{R=h\wedge h \,C} =0 \,.
\ee
Being rather nontrivial,
this condition will be shown to admit a solution linking
the normalization coefficients in front of the free higher
spin action functionals for different fields.

Let us now sketch the general procedure
for the search of the $AdS_5$ higher spin action.
In accordance with (\ref{S3}) we shall look for a
Lagrangian 5-form  bilinear in the higher spin curvatures
with some 1-form $U_{\Omega\Lambda}$ built from the higher spin
gauge 1-forms. As no useful extension of the compensator
formalism to the full higher spin algebra is known so far,
we use a mixed approach with the frame field
$E^{\ga\gb}$ built from the compensator
$V^{\ga\gb}$ and the gravitational fields associated with the
$AdS_5$ subalgebra $su(2,2) \subset$\hsa. In addition,
some explicit dependence on the higher spin gauge fields taking values
in \hsa $/su(2,2)$ will be allowed. Presumably, such an approach is a
 result of a partial gauge fixing in a full
compensator formalism in the $AdS_5$ higher spin theory to be developed.
Note that, perturbatively,
$E^{\ga\gb}$ contains the background gravitational field
and, therefore, is of the zero order, while the higher spin fields
are of the first order.

In our analysis the higher spin gauge fields will be allowed to
take values in some associative
(e.g., matrix) algebra $\go \to \go_I{}^J$.
The resulting ambiguity is equivalent to the ambiguity of
a particular choice of the Yang-Mills gauge algebra in the spin 1
sector.
The higher spin action will be formulated in terms of the trace
$tr$ in this matrix algebra (to be not confused with the trace
in the star product algebra). As a result, only cyclic permutations
of the matrix factors will be allowed under the trace operation.
Note that the
gravitational field is required to take values in the center of the matrix
algebra, being proportional to the unit matrix. For this reason, the
factors associated with the gravitational field
are usually written outside the trace.

Let us consider an  action of the form
\be
\label{stot}
S= S^E +S^\go\,,
\ee
where
\bee
\label{ans}
S^E =\half  \int_{M^5} \Big ( \ga E_{\ga\gb}
\frac{\partial^2}{\partial a_{1\ga} \partial a_{2\gb }}
&+&\gb E^{\ga\gb}
\frac{\partial^2}{\partial b_1^\ga \partial b_2^\gb }
+
\sum_{ij=1}^2 \gamma^{ij} E_{\ga}{}^{\gb}
\frac{\partial^2}{\partial a_{i\ga} \partial b_j^{\gb }}
 \Big )\wedge \nn\\
&{}&\ls\ls       tr(  R(a_1 ,b_1 |x )
\wedge     R(a_2 ,b_2 |x ) )\Big |_{a_1 = b_1 = a_2 =b_2 =0}\,
\eee
and
\bee
\label{shs}
S^\go =\half  \int_{M^5}  \tau tr(
 R(a_1 ,b_1 |x )
\wedge R(a_2 ,b_2 |x )\wedge \go (a_3 ,b_3 |x ))\Big |_{a_i = b_j =0}\,.
\eee
Here the coefficients $\alpha,\beta,\gamma^{ij}$ and $\tau$
are some functions of the
Lorentz invariant combinations of derivatives with respect to the spinor
variables $a_{i\ga}$ and $b^\ga_j$,
\be
\label{abc}
\bar{a}_{ij} =
V_{\ga\gb}\frac{\partial^2}{\partial a_{i\ga} \partial a_{j\gb }}\,,\qquad
\bar{b}_{ij}= V^{\ga\gb}\frac{\partial^2}{\partial b_{i}^{\ga}
\partial b_{j}^{\gb }}\,,\qquad \bar{c}_{ij} =
\frac{\partial^2}{\partial a_{i\ga} \partial b_j^{\ga }}\,
\ee
($i,j = 1,2$ for (\ref{ans}) and $ 1,2,3$ for (\ref{shs})).
Functions $\alpha,\beta,\gamma^{ij}$ and $\tau$
parametrize the ambiguity in all possible contractions of indices
of the component higher spin fields and curvatures.
Note that  the gravitational field is not allowed to appear
among the components of the connection $\go$ that enters
explicitly the action (\ref{shs}). Instead, all terms with the
gravitational field in front of the curvature terms are collected
in the action $S^E$ (\ref{ans}). With this convention,
$S^E$ contributes both to the quadratic and to the cubic parts of
the action while $S^\go$ only contributes to the interaction part of
the action.

Below we show that there exists a consistent cubic
higher-spin-gravitational interaction
for $S^\go =0$.
Since the aim of this paper is to show that at least some
consistent higher-spin-gravitational
interaction exists in $AdS_5$, we shall mostly focus on this particular
case. Note that it is anyway  hard to judge
on a full structure of the theory
from the perspective of the cubic interactions.
Indeed, at the cubic level one can switch out interactions among
any three elementary (i.e. irreducible at the free field level)
fields without spoiling the consistency at this order. This is most
obvious from the Noether coupling interpretation of the cubic
interactions: setting to zero some of the fields is always consistent
with the conservation  of currents. It is plausible to speculate
that the action $S^E$ accounts the terms relevant to the
higher-spin-gravitational interaction but may miss some other higher spin
interactions described by the action $S^\go$. Indeed, as a by product of
the consideration below we shall give
an example of a consistent higher spin interactions $S^\go$. Note that
even writing down all terms of the form (\ref{stot}) there is
little chance
to have a fully consistent theory beyond the cubic order
without introducing more dynamical fields because,
as we know from the $4d$ example \cite {Ann}
(see also \cite{R1,gol}), some lower spin fields
(e.g. spin 1 and spin 0) have to be added.
Note that the actions for spin 1 and spin 0 massless fields do not admit
a formulation in the form (\ref{S3}). To simplify the presentation
we will assume in this paper that these fields are set to zero, that
is a consistent procedure at the cubic order. By analogy with
the $4d$ case \cite{FV2} we expect that an extension of the
results of this paper to the full system with the lower spin fields
will cause no problem. Let us note that an appropriate reformulation of
the Lagrangian spin 0 free field dynamics was developed in \cite{SVsc}.

Let us now focus on the structure of the action $S^E$.
The ambiguity in the coefficient functions
$\ga,\gb,\gamma_{ij}$ can be
restricted by not allowing a contraction of the
both of indices of $ E^{\ga}{}_{\gb}$ with the same curvature.
Another restriction we impose is that a total number of derivatives
in $a_1$ and $b_1$ is equal to the number of derivatives
in $a_2$ and $b_2$, i.e. the terms resulting from the
products of the polynomials of different
powers in $R(a_1 , b_1)$ and $R(a_2 , b_2)$
are required to vanish. (The most important argument
for this ansatz is, of course,
 that it will be proved to work.) We therefore
consider the action of the form (\ref{ans}) with the coefficients
$\gga^{11}=\gga^{22}=0$. Taking into account that the higher spin gauge fields
and curvatures carry equal numbers of lower and upper indices, i.e.
$R(\mu a , \mu^{-1} b )=R( a , b ) $, the appropriate ansatz is
\be
\label{ans1}
S^E = \half A_{\ga,\gb,\gga}^E (R,R)\,,
\ee
where the symmetric bilinear  $A_{\ga,\gb,\gga}^E (f,g)= A_{\ga,\gb,\gga}^E
(g,f)$ is defined for any 2-forms $f$ and $g$ as
\bee \label{bform}
A^E_{\ga,\gb,\gga} (f,g)\!\! &=&\!\! \int_{M^5} \Big ( \ga(p,q,t ) E_{\ga\gb}
\frac{\partial^2}{\partial a_{1\ga}\partial a_{2\gb }} \bar{b}_{12} +\gb(p,q,t)
E^{\ga\gb} \frac{\partial^2}{\partial b_1^\ga \partial b_2^\gb } \bar{a}_{12}
\nn\\ &+& \gamma (p,q,t) \Big (E_{\ga}{}^{\gb}
\frac{\partial^2}{\partial a_{1\ga} \partial b_2^{\gb }} \bar{c}_{21} -
E^{\ga}{}_{\gb} \frac{\partial^2}{\partial b_1^{\ga} \partial a_{2\gb }}
\bar{c}_{12}\Big )\Big )\nn\\ &{}&\wedge
tr ( f(a_1 ,b_1 |x )\wedge g(a_2 ,b_2 |x ) )\Big |_{a_i=b_j=0}\,
\eee
where we use notations
\bee
\label{pqt}
p = \bar{a}_{12} \bar{b}_{12}
\qquad
q= \bar{c}_{12}\bar{c}_{21} \,,\qquad
t= \bar{c}_{11}\bar{c}_{22} \,.
\eee
The labels  $\ga,\gb$, $\gga$ and $E$ in
$A_{\ga,\gb,\gga}^E (f,g)$ refer to the functions
$\ga(p,q,t),\gb(p,q,t)$, $\gga(p,q,t)$ and the frame field
$E^{\ga\gb}$  that fix a particular form of the bilinear form.
Sometimes we will write $A(f,g) $  instead of
$A^E_{\ga,\gb,\gga}(f,g)$.

As explained in section \ref{Higher Spin Extension},
nonlinear actions of this  form  cannot have the
invariant trace property, i.e.
$A(a*f,g) \neq A(f,g*a) $ for generic  $a,f,g \in$\hsa.
One can require however a weaker condition
\be
\label{cinv}
A (N*f,g)= A (f, g *N)\,,
\ee
where $f$ and $g$ are any
elements satisfying
$
f*N =N*f\,,\quad g*N = N*g\,.
$
 {}From (\ref{cinv}) it follows that
\be
\label{Cinv}
A (\phi (N) *f,g)=A (f, g *\phi (N))\,.
\ee
We will refer to the property (\ref{cinv}) as the {\it $C-$invariance
condition}. It will play the key role in the analysis of the
invariance of the cubic action in section
\ref{Cubic Interactions}. The explicit form of the restrictions on the
coefficients $\ga,\gb,\gga$ due to (\ref{cinv}) is given in section
\ref{Quadratic Action}.

The main steps of the rest of the analysis are as follows.
First we analyze the quadratic part of the action choosing the
functions $\ga,\gb$ and $\gga$ to guarantee that the free action $S_2$
describes a sum of compatible with
unitarity free field actions for the set of the
higher spin fields associated with the higher spin algebra \hsa.
This is equivalent to the two conditions. First, the
{\it extra field decoupling condition} requires the
variation of the quadratic action with respect to the extra fields
to vanish. Second, the quadratic action should decompose into
infinite sum of free actions for the
different copies of fields of the same spin associated with the
spinor traces  as discussed below (\ref{lgf}). This is referred
to as the {\it factorization condition}.

Note that at the free field level
there is an ambiguity in the coefficients
$\ga(p,q,t)$ and $\gb(p,q,t)$ due to the freedom in adding the total
derivative terms
\bee
\label{totder}
\delta S^2 &=& \half  \int_{M^5}d  \Big (  \Phi (p,q,t )
tr ( R_1(a_1 ,b_1 |x )\wedge R_1 (a_2 ,b_2 |x ) )
|_{a_1 = b_1 = a_2 =b_2 =0}
\Big )\nn\\
&=&
\half  \int_{M^5} \frac{\partial \Phi (p,q,t)}{\partial p}\Big (
h^{\ga\gb}
\frac{\partial^2}{\partial b_1^\ga \partial b_2^\gb }
\bar{a}_{12}
-
h_{\ga\gb}
\frac{\partial^2}{\partial a_{1\ga}\partial a_{2\gb }}
\bar{b}_{12}
\Big )\nn\\
&{}&\wedge tr (R_1 (a_1 ,b_1 |x )
\wedge R_1 (a_2 ,b_2 |x )) \Big |_{a_i = b_j =0}\,,
\eee
where, using the manifest $su(2,2)$ covariance of our formalism,
the differential $d$ in the first line is replaced by the background
$su(2,2)$ derivative, and  the definition of the background
frame  field (\ref{h}) has been taken into account
along with the Bianchi identities $D_0 (R_1)=0$.  As a result, the variation
of the coefficients
\be
\label{de}
\delta \ga (p,q,t) = \epsilon (p,q,t)\,,\qquad
\delta \gb (p,q,t) = -\epsilon (p,q,t)
\ee
does not affect the physical content of the quadratic
action, i.e. only the combination $\ga (p,q,t)+\gb (p,q,t)$ has invariant
meaning at the free field level.
Modulo the  ambiguity (\ref{de}) the {\it
factorization condition} along
with the {\it extra field decoupling condition} fix the
functions $\ga ,\gb ,\gga $ up to an arbitrary function
parametrizing the ambiguity in the normalization coefficients
in front of the individual free actions. The proof of this
fact is the content of section \ref{Quadratic Action}.

In the analysis of the cubic interactions, there are two types of terms
to be taken into account.
Terms of the first type result from the gauge transformations
of the gravitational fields and the compensator $V^{\ga\gb}$
that contribute into the factors in front of the higher spin curvatures
in the action $(\ref{ans})$. The proof of the respective invariances
goes the same way as in the example of gravity considered in section
\ref{$AdS_d$ Gravity with Compensator} as it is based entirely on the
explicit $su(2,2)$ covariance and  invariance of the whole
setting under diffemorphisms (recall that the additional
invariance (\ref{adtr}) was identified in section
\ref{$AdS_d$ Gravity with Compensator} with a mixture of the
diffeomorphisms and  $su(2,2)$ gauge transformations.) Also, one
has to take into account that the higher spin
gauge transformation of the gravitational fields is at least linear in the
dynamical fields and therefore has to be discarded in the analysis of $\go^2
\gvep $ type terms under consideration.

The nontrivial terms of the second type originate from the variation
(\ref{Rtr}) of the higher spin curvatures. According to (\ref{ncon})
the problem is to find such functions
$\ga ,\gb$ and $\gamma$ that
\bee
\label{SEtr}
\delta^g S^E (R,R)
 \Big |_{E=h, R = h\wedge h C}
\equiv
 A^h_{\ga,\gb,\gga} (R,[R,\gvep ]_*)
 \Big |_{R = h\wedge h C} =0\,
\eee
for an arbitrary gauge parameter $\gvep (a,b |x )$. As shown
in section \ref{Cubic Interactions} this condition fixes
the coefficients in the form
\be
\label{gagbc}
\ga (p,q,t) + \gb (p,q,t)  =  \varphi_0  \sum_{m,n =0}^\infty
(-1)^{m+n} \f{m+1}{2^{2(m+n+1)} (m+n+2)! m! (n+1)!} p^n q^m\,,
\ee
\be
\label{ggac}
\gga (p,q,t)  =
\gga (p+q) \,,\qquad
\gga (p)=
\varphi_0  \sum_{m =0}^\infty
(-1)^{m+1} \f{1}{2^{2m+3} (m+2)! m!} p^m\,,
\ee
where $\varphi_0 $ is an arbitrary normalization factor to be
identified with the (appropriately normalized in terms of the
cosmological constant) gravitational coupling constant.
Let us note that the sign factors in the coefficients
(\ref{gagbc}) and (\ref{ggac}) distinguish between the polynomials
of the oscillators $a_\ga$ and $b^\gb$ of degree $4p$ and $4p+2$.
Together with the signs  due to the factors of $i$ in the normalization
coefficients (\ref{irf}) this implies that fields of equal
spins contribute to the quadratic action with the same sign. The
fields of even and odd spins contribute with opposite signs.

As a result,  the condition that
cubic $AdS_5$ higher spin action possesses higher spin gauge
symmetries fixes uniquely the relative coefficients in front
of the free actions of fields of all spins in a way compatible
with unitarity.
Note that the analysis of the interactions does not fix the
ambiguity (\ref{de}). Taking into account (\ref{totder})
along with the full Bianchi identities, one observes that the
ambiguity in $\ga-\gb$ is equivalent to the ambiguity in the
interaction terms $S^\go$ of the form
\bee
\label{sgo}
S^\go &=& \half  \int_{M^5}   \Phi (p,q,t )
tr \Big ([\go_1 , R]_* (a_1 ,b_1 |x ) \wedge R (a_2 ,b_2 |x )\nn\\
&+&
R (a_1 ,b_1 |x ) \wedge [\go_1 , R]_*  (a_2 ,b_2 |x ) \Big )
\Big |_{a_i = b_j =0}\,
\eee
parametrized by an arbitrary function $\Phi (p,q,t)$.

\subsection{Quadratic Action}
\label{Quadratic Action}

The free field part $S_2$ of the action $S$ is obtained
from (\ref{ans1})  by the
substitution of
the linearized curvatures (\ref{lin}) instead of $R$ and
$h^\ga{}_\gb$  instead of $E^\ga{}_\gb$.
The resulting action
is manifestly invariant under the linearized transformations
(\ref{lintr1}) because the linearized curvatures are invariant.
We want the free action to be a sum of
actions for the irreducible higher spin
fields we are working with. This requirement is not completely
trivial  because of the infinite degeneracy of the algebra
due to the traces. The {\it factorization condition} requires
\be
\label{S2sum}
S_2 =\sum_{s,n=0}^\infty S^{s,n}_{2} (\go_n^s  )\,,
\ee
i.e., the terms  containing products of the fields $ \go_n^s$
and $ \go_m^s $ should all vanish for $n\neq m$. As follows from
(\ref{irf}), (\ref{T0s}) along with (\ref{sl2inv}) this
is true if
\be
\label{kfact}
A^E_{\ga,\gb,\gga} ( f, (T^+ )^k g ) =
A^E_{\ga_k,\gb_k,\gga_k} ((T^- )^k f, g )\,,\qquad \forall k
\ee
for some $\ga_k,\gb_k$ and $\gga_k$.

An elementary computation shows that
\bee
A^E_{\ga,\gb,\gga}(f, T^+ g) &=&
A^E_{\ga_1,\gb_1,\gga_1}(T^- f,  g)\nn\\
&{}&\ls\ls\ls\ls\ls\ls\ls\ls+
   \int_{M^5}  Q (p,q,t ) E_{\ga}{}^{\gb}
\frac{\partial^2}{\partial a_{1\ga}\partial b_1^{\gb }}
\wedge tr \Big ( f(a_1 ,b_1 |x )\wedge
g (a_2 ,b_2 |x ) \Big ) \Big |_{a_i = b_j =0} \,,
\eee
where
\be
\label{fact0}
Q= \left ( 1+ p\frac{\partial}{\partial p} \right )
\Big (\ga(p,q,t) +\gb(p,q,t)\Big )
+2 \left ( 1+ q\frac{\partial}{\partial q} \right ) \gga(p,q,t) \,
\ee
and
\bee
\label{A}
{\ga_1} (p,q,t) \!\!&=&\!\!4 \left (
(2+p\frac{\partial}{\partial p} )
 \frac{\partial}{\partial p } +
(1+q\frac{\partial}{\partial q} )
 \frac{\partial}{\partial q } +\Big (2p\frac{\partial}{\partial p}
+2q\frac{\partial}{\partial q} +t\frac{\partial}{\partial t} +6 \Big )
\frac{\partial}{\partial t} \right )\nn\\
&{}&\times \ga (p,q,t )\,,
\eee
\bee
\label{B}
{\gb_1}(p,q,t) \!\!&=&\!\! 4 \left (
(2+p\frac{\partial}{\partial p} ) \frac{\partial}{\partial p } +
(1+q\frac{\partial}{\partial q} )
 \frac{\partial}{\partial q } +\Big ( 2p\frac{\partial}{\partial p}
+2q\frac{\partial}{\partial q} +t\frac{\partial}{\partial t} +6 \Big )
\frac{\partial}{\partial t} \right )\nn\\
&{}&\times \gb (p,q,t )\,,
\eee
\bee
\label{C}
{\gga_1}(p,q,t) \!\!&=&\!\! 4 \left (
(1+p\frac{\partial}{\partial p} )
 \frac{\partial}{\partial p } +
(2+q\frac{\partial}{\partial q} )
 \frac{\partial}{\partial q } +\Big (2p\frac{\partial}{\partial p}
+2q\frac{\partial}{\partial q} +t\frac{\partial}{\partial t} +6 \Big)
\frac{\partial}{\partial t} \right )\nn\\
&{}&\times \gga (p,q,t )\,.
\eee
The {\it factorization condition} therefore requires
\be
\label{fact1}
Q=\left ( 1+ p\frac{\partial}{\partial p} \right ) (\ga(p,q,t)
+\gb (p,q,t))
+2 \left ( 1+ q\frac{\partial}{\partial q} \right ) \gga(p,q,t) =0\,.
\ee
Then one observes that from (\ref{fact1}) it follows
that the same is true for the coefficients $\ga_1$, $\gb_1$ and $\gga_1$
(\ref{A})-(\ref{C}), and, therefore, (\ref{fact1}) guarantees
(\ref{kfact}) for all $k$. In the sequel, the {\it factorization
condition} (\ref{fact1}) is required to be true. Since the operator
$\Big ( 1+ q\frac{\partial}{\partial q} \Big )$ is invertible,
it allows to express $\gamma$  in terms of $\ga$ and $\gb$.

Let us now analyze the {\it {} $C-$invariance condition}
(\ref{cinv}).
Taking into account (\ref{NT}) along with the factorization condition
(\ref{kfact}), it amounts to
\be
A^E_{\ga,\gb,\gga} (T^- f , g) + A^E_{\ga_1,\gb_1,\gga_1} (T^- f , g) =
A^E_{\ga,\gb,\gga} (f , T^- g) + A^E_{\ga_1,\gb_1,\gga_1} ( f , T^- g)\,.
\ee
Obviously, this is true iff
\be
\label{01}
A^E_{\ga,\gb,\gga} ( f , g) =- A^E_{\ga_1,\gb_1,\gga_1} ( f , g)\,,
\ee
i.e.
\be
\label{CINV}
\ga (p,q,t)  =- \ga_1 (p,q,t)\,,\quad
\gb (p,q,t)  = -\gb_1 (p,q,t)\,,\quad
\gga (p,q,t)  =- \gga_1 (p,q,t)\,.
\ee
This is equivalent to the requirement that the operators $T^+$ and $-T^-$
are conjugated with respect to the bilinear form
$A^E_{\ga,\gb,\gga} ( f , g)$
\be
\label{conjT}
A^E_{\ga,\gb,\gga} ( T^\pm f , g)= -
A^E_{\ga,\gb,\gga} (  f ,T^\mp  g)\,.
\ee
Let us note that the original ansatz for the bilinear form
(\ref{bform}) satisfies
\be
\label{conjT0}
A^E_{\ga,\gb,\gga} ( T^0 f , g)=
A^E_{\ga,\gb,\gga} (  f ,T^0    g)\,.
\ee

{}From (\ref{A})-(\ref{C}) it is clear that (\ref{CINV})
reconstructs the dependence of $\ga (p,q,t)$,
$\gb (p,q,t)$ and $\gga (p,q,t)$  on $t$ in terms of the
``initial data" $\ga (p,q,0)$, $\gb (p,q,0)$ and $\gga (p,q,0)$.

With the help of (\ref{conjT}) along with  (\ref{sl2inv})
it is elementary to compute the relative coefficients of the
actions for the
different copies of fields in the decomposition (\ref{irf}),
(\ref{hw}). The coefficients (\ref{coe} ) are fixed so that the
linearized actions have the same normalization for different copies of
the higher spin fields parametrized by the label $n$
\be
\label{S2summ}
S_2 =\sum_{s,n=0}^\infty S^{s}_{2} (\go_n^s  )\,.
\ee
In  the
linearized approximation it is therefore enough to analyze the situation
for
any fixed $n$. We confine ourselves to the case of
$\go = \go_0$ assuming in the rest of this section that
\be
\label{T-go}
T^- \go = 0\,.
\ee

Let us now consider the
{\it extra field decoupling condition}.
 Since the generic variation of the linearized
higher spin curvature is $\delta R_1 = D_0 \delta \go\,,$ where
$D_0$ is the $AdS_5$ background covariant derivative and
because the action is formulated in the $AdS_5$ covariant way
with the aid  of the compensator field $V^{\ga\gb}$,
integrating by parts one obtains for the generic variation of $S_2$
\bee
\delta S_2 =  \int_{M^5}D_0 \Big (&{}& \ls \ls \ga (p,q,0 ) h_{\ga\gb}
\frac{\partial^2}{\partial a_{1\ga}\partial a_{2\gb }}
\bar{b}_{12}
+\gb (p,q,0 ) h^{\ga\gb}
\frac{\partial^2}{\partial b_1^\ga \partial b_2^\gb }
\bar{a}_{12}\nn\\
&+&
\gga (p,q,0) \Big ( h_{\ga}{}^{\gb}
\frac{\partial^2}{\partial a_{1\ga} \partial b_2^{\gb }} \bar{c}_{21}
-h^{\ga}{}_{\gb}
\frac{\partial^2}{\partial b_1^{\ga} \partial a_{2\gb }}\bar{c}_{12}\Big )
\Big )
\nn\\
&{}&\wedge tr ( \delta \go(a_1 ,b_1 |x )
\wedge R_1 (a_2 ,b_2 |x ))\Big |_{a_i = b_j =0}\,,
\eee
where it is taken into account that the $t-$dependent terms
trivialize as a consequence of (\ref{T-go}).
The derivative $D_0$
produces the frame field every time it meets
the compensator. (Recall that
$D_0 (h ^{\ga\gb})=0 $ because $D_0^2 = R_0 = 0$.)
Taking into account
(\ref{id1}) and (\ref{T-go}), one finds
\bee
\label{deltaS2}
\delta S_2 =  \half \int_{M^5} &{}& \ls \rho (p,q) \Big (
\frac{\partial^2}{\partial a_{1\ga}\partial b_2^{\gb }}
\bar{c}_{21}
+
\frac{\partial^2}{\partial a_{2\ga}\partial b_1^{\gb }}
\bar{c}_{12}
\Big )\nn\\
&{}& H_{2\,\ga}{}^{\gb} \wedge tr (\delta \go(a_1 ,b_1 |x )
\wedge
R_1 (a_2 ,b_2 |x )\Big |_{a_i = b_j =0} \,,
\eee
where
\be
\label{E}
\rho(p,q)  = \Big (1 +p\f{\p}{\p p} \Big )
\Big (\ga (p,q,0) +\gb (p,q,0)  -2 \gga (p,q,0)\Big )\,.
\ee

According to (\ref{sviden})
the extra fields are associated with the
multispinors described by the two-row Young diagrams
of the Lorentz algebra  having at least four more cells in the upper
row than in the lower one.  As follows from (\ref{lgf}),
generic variation sharing this property has the form
\be
\label{extvar1}
\delta \go^{ex}(a , b ) =(S^+ )^2\xi (a , b ) \,.
\ee
To guarantee $(N_a - N_b  )\delta \go^{ex}(a , b ) =0$,
the infinitesimal $\xi (a , b )$ is required
to satisfy
$
(N_a - N_b -4 ) \xi (a , b ) =0.
$

To derive the restriction on the coefficients imposed by the
requirement that the extra fields do not contribute to the
variation one observes, first, that
\be
 \Big [ S^+_1 ,q \Big ]=
-\Big [S^+_1  ,p \Big ]=u\,,\qquad [S^+_1 , u] =0\,,
\ee
where
\be
S^+_1 = b_{1\gb} \f{\p}{\p a_{1\gb} }\,,\qquad
u= \bar{a}_{12}\bar{c}_{12}
\ee
and, second, that the double commutator of $S^+_1$
to the differential operator next to $\rho (p,q)$
in (\ref{deltaS2}) is zero.
As a result, substituting (\ref{extvar1}) into
(\ref{deltaS2}) one finds that the corresponding
variation of the action vanishes provided that
\be
(\frac{\partial}{\partial p }-
\frac{\partial}{\partial q } ) \rho( p,q) =0\,.
\ee
Therefore the {\it extra field decoupling condition} requires
\be
\label{ext}
\rho (p,q ) = \rho (p+q )\,.
\ee

{}From the { \it factorization condition} (\ref{fact1})
and (\ref{E}), (\ref{ext}) it follows that
\be
\label{g+}
\gga (p,q,0) = \gga (p+q,0)\,,\qquad \rho(r) =
-2 (r\frac{\partial}{\partial r}+2) \gga(r,0)\,.
\ee
The function $\ga (p,q,0) +\gb (p,q,0)$ is fixed in terms of $\gga (p+q,0)$
by the {\it factorization condition} (\ref{fact1})
\be
\label{gagga}
\ga (p,q,0) +\gb (p,q,0) = -2 \int_0^1 du \Big ( 1 + q \f{\p}{\p q}
\Big ) \gga (up +q ) \,.
\ee

The function of one
variable $\gga (p+q,0)$ parametrizes the leftover ambiguity
in the coefficients (discarding the trivial
ambiguity (\ref{de})) associated with
the ambiguity in the coefficients in front of
the free actions of fields with different spins.
Indeed, the total homogeneity degree in the variables $p$ and $q$,
 telling us how many pairs of indices are contracted,
equals to $s-1$. Clearly, this ambiguity
cannot be fixed from the analysis of the free action.

\subsection{Cubic Interactions}
\label{Cubic Interactions}

Let us now analyze the {\it on-mass-shell invariance condition}
 (\ref{ncon}) to prove the existence of
a nonlinear deformation of the higher spin gauge transformations
that leaves the cubic part of the action $S=S^E$ invariant to the
order $\go^2 \gvep$. As explained in the beginning of this section
this condition amounts to (\ref{SEtr}).
Taking into account (\ref{ircu}),
our aim is to prove that there exist such coefficient functions
$\ga,\gb$ and $\gga$ satisfying the {\it {} C-invariance
condition}, {\it factorization condition} and {\it
extra field decoupling condition} that
\be
\label{sinvmn}
\sum_{mn} A^h_{\ga,\gb,\gga}
\left ((T^+)^m v_m (T^0 ) R_{1,m} (a ,b )\Big |_{m.s.} ,
[\gvep , (T^+)^n v_n (T^0 )  R_{1,n} (a ,b )\Big |_{m.s.} \,]_*
\right ) =0
\ee
for any gauge parameter $\gvep \in$\hsa and arbitrary
Weyl tensors $C_n (a)$ in the  spinor form (\ref{scomt})
of the First On-Mass-Shell Theorem.

To this end,
one first of all observes that the dependence of $v_n (T^0 )$ on $T^0$
can be absorbed into (spin-dependent) rescalings
of the Weyl tensors $C_n (a)$ which are treated as arbitrary
field variables in this consideration. As a result it is enough to
prove (\ref{sinvmn}) for arbitrary constant coefficients $v_n$.

Now let us show that, once (\ref{sinvmn})  is valid for  $m=n=0$,
it is automatically true for all other values of $m$ and $n$ as a
consequence of the {\it {} $C-$invariance condition}.
Indeed, suppose that (\ref{sinvmn}) is true for
$m_0 \geq m \geq 0$, $n_0 \geq n \geq 0$. Consider the term with
$m= m_0 +1$. Then, from (\ref{NT}) it follows
\be
(T^+)^{m_0 +1} R_{1,m_{0+1}} (a ,b )=
N*( (T^+)^{m_0} R_{1,m_{0+1}} (a ,b )) +
T^-  (T^+)^{m_0} R_{1,m_{0+1}} (a ,b )\,.
\ee
The term containing $T^-  $ gives zero contribution
by the induction assumption since,
taking into account (\ref{T-R}), $T^-$ decreases a number of
$T^+$.
By virtue of the {\it {}
$C-$invariance condition} (\ref{cinv}) along with the fact that
$N$ belongs to the center of \hsa so that
\be
\label{Nfg}
(N*f)*g = f* (N*g) \,,
\ee
the term containing the star product with $N$ equals to
\be
A^h_{\ga,\gb,\gga}
\left ((T^+)^{m_0}v_{m_0} (T^0 )   R_{1,m_0 } (a ,b )\Big |_{m.s.} ,
[N*\gvep , (T^+)^{n_0} v_{n_0} (T^0 )  R_{1,n_0} (a ,b )\Big |_{m.s.}
\,]_* \right )
\ee
which is zero by the induction assumption
valid for any $\gvep$. Analogously,
one performs induction $n_0 \to n_0 +1$ with the aid of (\ref{Nfg}).

Thus, it suffices to find the coefficients satisfying the
{\it {} $C-$invariance condition} for $R={\cal R}\equiv R_{1,0}$.
In other words one has to prove that
\be
\label{3inv}
A^h_{\ga,\gb,\gga}
\left ( {\cal R}  ,
[\gvep ,  {\cal R} \,]_*
\right ) =0
\ee
for
\be
\label{cR}
{\cal R} (a ,b ) =
H_{2\ga}{}^\gb \f{\p^2}{\p a_\ga \p b^\gb }
Res_\mu (C (\mu a + \mu^{-1} b) )\,.
\ee

Note that because  $T^- ({\cal R}) = 0$ the terms containing $\bar{c}_{11}$
(\ref{abc}) and, therefore, $t$ (\ref{pqt}) does not contribute into
the condition (\ref{3inv}).

Using the differential (Moyal)  form of the star product (\ref{prod})
one finds
\bee
\label{SEtrc}
A^h_{\ga,\gb,\gga} ( f \!\!\! &,&\! \!\!\eta  * \gvep ) =
\int_{M^5} e^{\half (\bar{c}_{23} - \bar{c}_{32} )}
 \Big \{(\bar{b}_{12} +\bar{b}_{13} )
 h_{\ga\gb}
\Big (\frac{\partial^2}{\partial a_{1\ga}\partial a_{2\gb }}+
\frac{\partial^2}{\partial a_{1\ga}\partial a_{3\gb }} \Big )\nn\\
&{}&\times
\ga \Big ((\bar{a}_{12} +\bar{a}_{13})(\bar{b}_{12} +\bar{b}_{13})\,,
(\bar{c}_{12} +\bar{c}_{13})(\bar{c}_{21} +\bar{c}_{31}),0\Big )\nn\\
 &{}& \ls \ls + (\bar{a}_{12} +\bar{a}_{13} )
 h^{\ga\gb}
\Big (\frac{\partial^2}{\partial b_1^{\ga}\partial b_2^{\gb }}+
\frac{\partial^2}{\partial b_{1}^\ga\partial b_3^{\gb }} \Big )\nn\\
&{}&\times
\gb\Big ((\bar{a}_{12} +\bar{a}_{13})(\bar{b}_{12} +\bar{b}_{13})\,,
(\bar{c}_{12} +\bar{c}_{13})(\bar{c}_{21} +\bar{c}_{31}),0\Big )\nn\\
 &{}& \ls \ls \ls \ls +\Big (
(\bar{c}_{21} +\bar{c}_{31} )
 h_{\ga}{}^{\gb}
\Big (\frac{\partial^2}{\partial a_{1\ga}\partial b_2^{\gb }}+
\frac{\partial^2}{\partial a_{1\ga}\partial b_3^{\gb }} \Big )
-(\bar{c}_{12} +\bar{c}_{13} )
 h^{\ga}{}_{\gb}
\Big (\frac{\partial^2}{\partial b_1^{\ga}\partial a_{2\gb }}+
\frac{\partial^2}{\partial b_1^{\ga}\partial a_{3\gb }} \Big )\Big )\nn\\
&{}&\times
\gga\Big ((\bar{a}_{12} +\bar{a}_{13})(\bar{b}_{12} +\bar{b}_{13})\,,
(\bar{c}_{12} +\bar{c}_{13})(\bar{c}_{21} +\bar{c}_{31}),0\Big )\Big \}
\nn\\
&{}& tr \Big ( f(a_1 ,b_1 ) \eta (a_2 ,b_2 )
\gvep (a_3 ,b_3 )\Big ) \Big |_{a_i = b_j =0} \,
\eee
provided that $T^- f =0$. Let us consider
$A^h_{\ga,\gb,\gga} ( {\cal R}  ,{\cal R}  * \gvep )$.
Rewriting (\ref{cR}) as
\be
\label{ECMT}
{\cal R} (a_i ,b_i ) = Res_{\mu_i} e^{\mu_i a_{i\ga} \f{\p}{\p c_{i\ga}} +
\mu_i^{-1} b_i^{\ga} \f{\p}{\p c_i^{\ga}}}
H^{\ga\gb}\f{\p^2}{\p c_i^\ga \p c_i^\gb} C(c_i)\Big |_{c_i=0}\,,
\ee
and using notation
\be
\label{uv} \bar{k} = \f{\p^2}{\p c_{1\ga} \p c^\ga_2}\,,\qquad
\bar{u}_{i} = \f{\p^2}{\p c_i^{\ga} \p a_{3\ga}}\,,\qquad \bar{v}_{i} =
\f{\p^2}{\p c_{i\ga} \p b_{3}^\ga}\, \ee
along with the identities (\ref{E5}) and (\ref{id2}) applied to the background
fields, one finds \bee
A^h_{\ga,\gb,\gga} ( {\cal R} \!\! &,& \!\!{\cal R}  * \gvep ) =
B\int_{M^5}\bar{k}^2 H_5  Res_\mu \Big (
e^{\half (\mu \bar{v}_2 -
\mu^{-1} \bar{u}_2 )}
\varphi (Z) \Big )
\nn\\
&{}&
\times tr\Big (C(c_1 ) C(c_2 )
\gvep (a_3 , b_3 )\Big) \Big |_{c_1 = c_2 = a_3 = b_3 =0}\,,
\eee
where $B\neq 0$ is some numerical factor,
\be
\label{vp}
\varphi (Z) =  Z\Big (2 \gga (Z,-Z) - (\ga(Z,-Z) + \gb(Z,-Z) )\Big )
\ee
and
\be
Z= (\mu \bar{k} - \bar{u}_1 )(\mu^{-1} \bar{k} + \bar{v}_1 )\,.
\ee
(Note that the dependence on $\mu_1$ in the representation
(\ref{ECMT}) for the first factor of ${\cal R}$ cancels out while $\mu = \mu_2$
for the analogous representation in the factor of ${\cal R}$ in ${\cal R}
*\gvep$.)

Analogously, after recycling the product factors under
the matrix trace $tr$ and renumerating the spinor variables one obtains
\bee
A^h_{\ga,\gb,\gga} ( {\cal R} \!\! &,& \!\!\gvep * {\cal R}   ) =
B\int_{M^5}\bar{k}^2 H_5
Res_\mu \Big (  e^{-\half (\mu \bar{v}_1 - \mu^{-1} \bar{u}_1)}
\varphi (Y) \Big )
\nn\\
&{}&
\times tr\Big (C(c_1 ) C(c_2 )
\gvep (a_3 , b_3 )\Big) \Big |_{c_1 = c_2 = a_3 = b_3 =0}\,,
\eee
where
\be
Y= (\mu \bar{k} + \bar{u}_2 )(\mu^{-1} \bar{k} - \bar{v}_2 )\,.
\ee
The problem therefore is to find such a function $\varphi (Y)$ that
\bee
\label{rav}
\bar{k}^2 Res_\mu \Big (
e^{\half (\mu \bar{v}_2 -
\mu^{-1} \bar{u}_2 )}
\varphi (Z) -
e^{-\half (\mu \bar{v}_1 - \mu^{-1} \bar{u}_1)}
\varphi (Y) \Big )\nn\\
\times tr\Big (C(c_1 ) C(c_2 )
\gvep (a_3 , b_3 )\Big) \Big |_{c_1 = c_2 = a_3 = b_3 =0}=0\,.
\eee

As a first guess let us try $\varphi (A B) = Res_\nu \Big (
e^{\half (\nu^{-1} A +\nu B) } \Big )$. Then the two terms
in brackets in (\ref{rav}) amount to
\be
\label{equ}
Res_{\mu , \nu} \Big (
e^{\half (\mu \bar{v}_2 - \mu^{-1} \bar{u}_2
+\nu^{-1}\mu \bar{k} - \nu^{-1}\bar{u}_1 +\nu\mu^{-1}\bar{k}
+\nu \bar{v}_1  )}
-e^{\half ( \mu^{-1}\bar{u}_1 - \mu \bar{v}_1 +\nu^{-1}\mu \bar{k}
+\nu^{-1} \bar{u}_2 +\nu \mu^{-1} \bar{k} - \nu \bar{v}_2 )}\Big )\,.
\ee
These cancel out upon substitution $\nu \leftrightarrow -\mu$.
However, this solution is not completely satisfactory because
the formula (\ref{vp}) requires $\varphi (Z)$ to vanish at $Z=0$
to have  analytic functions $\ga$, $\gb$, $\gga$.

The following comment is now in order. As discussed in the
beginning of this section, throughout this paper we only consider
interactions of the higher spin fields with spins $s\geq 2$.
{}From the perspective of
the First On-Mass-Shell Theorem in the form (\ref{ECMT}) this
implies that $C(c)$ starts from the fourth-order polynomials in the
spinor variables $c_\ga$, i.e.
\be
\label{conve}
C(0)=0\,,\qquad \f{\p^2}{\p c^\ga \p c^\gb}C(c) \Big|_{c^\ga =0} =0\,.
\ee
Since the factor $\bar{k}^2$ in (\ref{rav}) contains two
differentiations both in $c_1$ and in $c_2$,  (\ref{conve})
means that adding a constant to $\varphi $ does not affect
(\ref{rav}).
This allows one to cancel out a constant term in $\varphi$
by setting
\be
\varphi (A) =
\varphi_0 Res_\nu \Big (
e^{\half (\nu^{-1}  +\nu A) }        -1 \Big )\,.
\ee

As a result, the {\it on-mass-shell invariance condition} solves by
\bee
2 \gga (A,-A) - \ga(A,-A) - \gb(A,-A)
&\!\!=\!\!&
\varphi_0 A^{-1} \Big ( Res_\nu \Big (e^{\half (\nu^{-1}
+\nu A) } \Big ) -1 \Big )\nn\\
                      &{} &\!\!\ls\ls\ls\ls\ls\ls\!\!=
\half \varphi_0 \int^1_0 du \Big ( Res_\nu  \Big ( \nu e^{\half
(\nu^{-1}  +\nu u A) } \Big ) \Big ) \,.
\eee
Taking into account (\ref{g+}) and
(\ref{gagga}) this is solved by
\be
\label{intgga}
\gga (p) = \frac{1}{4} \varphi_0 \int^1_0 dv v
Res_\nu \Big ( \nu e^{\half (-\nu^{-1} +\nu vp )} \Big )\,
\ee
and
\be
\label{intga}
\ga (p,q,0)+\gb (p,q,0) = 2 \gga (p+q) - \half\varphi_0 \int^1_0 du
Res_\nu \Big ( \nu e^{\half (-\nu^{-1} +\nu
(u p +q) \Big )}\,.
\ee
Expansion of these expressions for $\gga (p)$ and
$\ga (p,q,0) +\gb (p,q,0)$
in the power series gives (\ref{gagbc}) and (\ref{ggac}).
With aid of these power series expansions one can see
that the following identities are true
\be
\label{to1}
\Big (p\f{\p^2 }{\p p^2} +3\f{\p}{\p p} +\f{1}{4} \Big ) \gga (p) = 0 \,,
\ee
\be
\label{to2}
\Big ( \Big ( 2 + p \f{\p}{\p p} \Big ) \f{\p}{\p p} +
\Big ( 1 + q \f{\p}{\p q} \Big ) \f{\p}{\p q} + \f {1}{4} \Big )
(\ga (p,q,0)+ \gb (p,q,0)) =0\,.
\ee
{}From (\ref{A}) - (\ref{C}) it follows then that the
{\it {} $C-$invariance condition} (\ref{CINV}) is satisfied with
\be
\ga (p,q,t) + \gb (p,q,t)= \ga (p,q,0)+\gb (p,q,0)\,,\qquad
\gga (p,q,t) = \gga (p,q,0)\,.
\ee

Thus it is shown that the coefficient functions
(\ref{gagbc}) and (\ref{ggac}) satisfy
the {\it factorization condition},
{\it {} $C-$invariance condition},
{\it extra field decoupling condition} and the
{\it on-mass-shell invariance
condition}. The  resulting  bilinear form (\ref{bform})
defines the action (\ref{ans1}) that properly describes the
higher spin dynamics both at the free field level and at the
level of cubic interactions. The leftover ambiguity in
the coefficients $\ga(p,q,t) +\gb (p,q,t)$ and $\gga (p,q,t)$
reduces to the overall factor $\varphi_0$ that encodes
the ambiguity in the gravitational  constant.

\section{Reduced Models}
\label{Maximally Reduced Model}

So far we discussed  the $5d$
higher spin algebra \hsa being the centralizer
of $N$ in the star product algebra. This algebra is not
simple as it  contains infinitely many ideals $I_{P(N)}$
spanned by the elements of the form $P(N)*f$ for any
$f \in $\hsa and any star-polynomial $P(N)$ \cite{FLA}. Considering the
quotient algebras \hsa$/I_{P(N)}$ is equivalent to ``imposing
operator constraints'' $P(N)=0$. In this section we
focus on the  algebra $hu_0 (1,0|8)$ that results from
 $P(N) = N$ and its further reduction $ho_0 (1,0|8)$.
The algebra  $hu_0 (1,0|8)$  corresponds to the system
of higher spin fields of all integer spins with every spin emerging
once. $ho_0 (1,0|8)$ is its reduction to the system of
all even spins. Both of these algebras are of interest
from the AdS/CFT perspective \cite{SSd,BHS} .

The explicit construction of $hu_0 (1,0)$ = \hsa$/N$ via factorization
is not particularly useful within the star product  setup
because  $N*$ is the second order differential
operator (\ref{N*}). A useful approach used in \cite{BHS} consisted
of taking projection by considering elements of the form $f*F$ where
$f$ was an element of \hsa while $F$ was  a certain Fock vacuum projector
satisfying $N*F =0$.
In fact, the left module over \hsa generated from
$F$ was shown in \cite{BHS} to describe $4d$ conformal fields. In this
construction, the factorization of \hsa to \hsa$/N$ was
automatic. The Fock vacuum $F$ was $4d$ Lorentz invariant and
had definite scaling dimension. It is not
invariant under the $AdS_5$ Lorentz algebra $o(4,1)$ however,
and therefore cannot be used for the $AdS_5$ bulk higher spin gauge
theory considered in this paper. On the other hand, from the perspective
of this paper the  Fock module construction is irrelevant. We therefore
relax the property that the projector is a Fock vacuum for certain
oscillators. Instead we shall look for a $su(2,2)$ invariant operator $M$
satisfying
\be
\label{NM}
N*M = M*N =0\,,
\ee
\be
\label{DM}
 D_0 (M) =0\,.
\ee

To satisfy (\ref{DM}) we choose a manifestly $su(2,2)$ covariant
ansatz $M=M(a_\ga b^\ga )$. For any
polynomial function $M$ this would imply that it is a star polynomial
of $N$. {}From (\ref{NM}) it is clear however that $M$ cannot be a star
product function of $N$. Nevertheless there is a unique
(up to a factor) analytic
solution for $M=M(a_\ga b^\ga )$ that solves (\ref{NM}). Indeed,
from (\ref{N*})  it follows that the condition (\ref{NM}) has the form
\be
\label{deq}
- x M(x) +
M^\prime (x) +\f{1}{4} x M^{\prime \prime} (x) =0\,,\qquad M^\prime = \f{\p
M}{\p x}\,.
\ee
This is solved by
\be
\label{M}
M (x) = \int_{-1}^1 dl (1-l^2) e^{2 l x }\,,
\ee
as one can easily see using $(2 x - \f{\p}{\p l} ) e^{2 l x }=0$ and
integrating by parts. Equivalently \be
M (x) = \left ( 1 -\f{1}{4} \f{\p^2 }{\p x ^2} \right ) \f{sh (2x)}{x}\,.  \ee
Note that $M(x)$ is even \be
\label{M-x} M(-x) = M (x)\,.
\ee

Having found the operator $M$ we can write the action
for the reduced system associated with $hu_0 (1,0|8)$ by replacing
the bilinear form in the action with
\be
\label{bformc}
A(f,g) \to A_0 (f,g)=  A(f,M*g)\,.
\ee
Note that $A_0 (f,g)$ is well-defined as a functional of
polynomial functions (or, formal power series) $f$ and $g$
for any entire function $M(a_\ga b^\ga )$ because, for
polynomial $f$ and $g$, only a finite number of terms in the
expansion of $M(a_\ga b^\ga )$ contributes.
The modification of the bilinear form according to
(\ref{bformc}) with any   $M(a_\ga b^\gb)$ leads to a new
invariant action (\ref{ans1}). The reason why
this ambiguity was not observed in our analysis
is that we have imposed the {\it factorization condition} in a
particular basis of the higher spin gauge fields, thus not allowing
the transition to the new bilinear form (\ref{bformc}).

All other conditions, namely, the {\it {} $C-$invariance condition},
{\it extra field decoupling condition} and the
{\it on-mass-shell invariance condition} remain valid for any
entire function $M(a_\ga b^\gb)$ inserted into the bilinear form.
The factorization condition is relaxed in this section. Note that
the {\it {} $C-$invariance condition} guarantees that the
bilinear form $A_0$ is symmetric
\be
A (f, M*g ) = A(f*M , g)\,.
\ee

Inserting a particular function $M(a_\ga b^\gb)$ (\ref{M})
we automatically reduce the system
to a smaller subset of fields being linear combinations
of the different copies of the fields emerged in the original \hsa
model. Namely, we can now require all fields in the expansion (\ref{gf})
to be traceless. In other words,
the  representatives of the quotient algebra \hsao are identified
with the elements $g$ satisfying the traceless condition
\be
\label{T-g}
T^- g =0\,.
\ee
Indeed, by virtue of (\ref{NT}) any polynomial $\tilde{g} (a,b)\in$\hsa is
equivalent to some $g$ satisfying (\ref{T-g}) modulo terms containing
star products with $N$ which trivialize when acting on $M$. The star product
$f*g$ of any two elements $f$ and $g$ satisfying the
tracelessness condition (\ref{T-g}) does not necessarily satisfy
the same condition, i.e. $T^- (f*g )\neq 0$ (otherwise the elements satisfying
(\ref{T-g}) would form a subalgebra rather than a quotient algebra).
However the difference is irrelevant inside the action
built with the help of the bilinear form $A_0$. In particular,
the higher spin field strength
\be
( d\go +\go \wedge *\go ) *M
\ee
is equivalent to that of the higher spin algebra $hu_0(1,0|8)$.

Thus, the action
\be
\label{ansred}
S^E_{red} = \half A^E_{\ga\gb\gga} (R *M , R )
\ee
leads to
a consistent free field description and cubic interactions
for the system of the higher spin fields associated with the
higher spin algebra $hu_0 (1,0|8)$. The resulting
system describes massless fields of all integer spins $s\geq 2$, every
spin emerges once. The further reduction to the subalgebra
$ho_0 (1,0|8)\subset hu_0 (1,0|8)$ associated with the subset of
even spins is now trivially obtained by setting to zero all fields of
odd integer spins. (For more details on the Lie algebraic definition
of the corresponding reduction we refer the reader to
\cite{SSd,BHS}). Note that according to the analysis of the
section \ref{Cubic Interactions} one can consider the
dynamical system with $n^2$ fields of each spin, taking values in
the matrix algebra $Mat_n$. This system corresponds to the
higher spin algebra $hu_0(n,0|8)$. Its reduction to $ho_0(n,0|8)$
describes higher spin fields of even spins in the symmetric
representation of $o(n)$ and odd spins in the adjoint representation
of $o(n)$. (Therefore, no odd spins for the case of $n=1$).

Note that the conclusions of this section sound
somewhat opposite to those of \cite{FL}
where it was claimed that the analogous reduction for the $4d$
conformal higher spin theories is inconsistent.

The following comment is now in order.
Since $M$ is a particular nonpolynomial (although entire)
function, one has to be careful in treating it as an element
of the star product algebra which in our setup is regarded
either as the algebra of polynomials or of formal power series.
Since $M( a_\ga b^\ga )$ is uniquely defined by the property
(\ref{NM}) and $M*M$ formally has the same property,  one might
expect that $M*M= m M$ with some numerical factor $m$. Once this would
be true, it would be possible to rescale $M$ to a projection operator.
However, this is not possible because the parameter $m$
turns out to be infinite. As this issue may be interesting beyond
the particular $5d$ problem studied in this paper,
let us consider the general case
with the indices $\ga ,\gb \ldots$ ranging from 1 to $2n$.
The equation (\ref{deq}) generalizes to
\be
\label{deqn}
- x M(x) +\half n M^\prime (x) +\f{1}{4} x M^{\prime \prime} (x) =0\,
\ee
with the solution
\be
\label{Mn}
M (x) = \int_{-1}^1 dl (1-l^2)^{n-1} e^{2 l x }\,.
\ee
An elementary computation then shows that
\be
\label{M*M}
(M*M) (x) = \int_{-1}^1 dl\int_{-1}^1 dl^\prime \int_{-1}^1 dk
\delta \Big (k-\f{l+l^\prime}{1+l l^\prime } \Big )
\f{(1-k^2 )^{n-1}}{(1+l l^{\prime})^2} \exp 2kx\,.
\ee
{}From this formula it follows that
the expression $M*M$ is ill-defined for any $n$
because the
factor $\f{1}{(1+l l^\prime )^2}$ gives rise to a divergency
at the boundary of the integration region.

Therefore,
one cannot treat elements like $M$ as elements of the star product
algebra. In particular, this concerns the construction suggested
in \cite{SSd} for the description of the $5d$ generating function for
the scalar massless field and higher spin Weyl tensors in terms of the
fields $\Phi (a,b|x)$ required to
satisfy the condition $N*\Phi=\Phi*N = 0$. {}From what is explained
in this section it is clear that
\be
\Phi = M* \phi = \phi *M \,,\qquad \forall \phi :\,[\phi ,N]_* = 0\,.
\ee
In particular $M$ itself belongs to this class. There is no problem at the
linearized level as far as the variables $\Phi$ are no multiplied, but it
is likely to be a problem at the interaction level.

Let us stress again that in the construction presented in this
section the appearance of $M$ in the action functional
causes no problem because it is only multiplied with polynomial
elements of the higher spin algebra and never with itself.

\section{Conclusion}
\label{Conclusion}

It is shown that $5d$ higher spin gauge theories
admit consistent higher spin interactions at the action level
at least in the cubic order and that,
in agreement with the conjecture
of \cite{vf} and the construction of $4d$ conformal higher spin
algebras of \cite{FLA},
$5d$ higher spin symmetry algebra admits a
natural realization in terms of certain star product algebras with
spinor generating elements. One difference compared to the $4d$
case is that the $5d$ higher spin algebra \hsa contains non-trivial
center freely generated by the  element $N$ (\ref{N}). As a
result, $5d$ higher spin algebra \hsa gives rise to the infinite sets
of fields of all spins. That every spin appears in infinite number of
copies makes the spectrum of the $5d$ higher spin theories
reminiscent of the string theory. On the other hand, we have shown
that the factorization of the algebra \hsa with respect to the
maximal ideal generated by $N$, that  gives rise to the
reduced higher spin algebra $hu_0(1,0|8)$ in which every integer
spin appears in one copy, admits consistent interactions as well.
The same is true for the further reduction the algebra $ho_0 (1,0|8)$
discussed in \cite{SSd}, that describes higher spin fields of even spins,
as well as for the matrix extensions $hu_0(n,0|8)$ and $ho_0(n,0|8)$
that describe either $n^2$ fields of every integer spin in the case
of $hu_0(n,0|8)$  or $\half n(n+1)$
fields of every even spin and $\half n(n-1)$
fields of every odd spin in the case of $ho_0(n,0|8)$.

The obtained results
are expected to admit a generalization to the supersymmetric
case. To this end one extends the set of oscillators $a_\ga$ and
$b^\gb$ with the set of Clifford elements $\phi_i$ and $\bar{\phi}^j$
($i,j = 1\ldots \N$) satisfying the commutation relations
\be
\{\phi_i , \phi_j \} =0\,,\qquad
\{\bar{\phi}^i , \bar{\phi}^j \} =0\,,\qquad
\{\phi_i , \bar{\phi}^j \} =\delta_i{}^j\,.
\ee
The supersymmetric extension of the $5d$ higher spin algebra
is then defined as the centralizer of
\be
N_\N = a_\ga b^\ga -\phi_i \bar{\phi}^i\,.
\ee
The $\N=1$ supersymmetric $5d$ higher spin theories will be
analyzed in \cite{AV}. An extension to $\N \geq 2$ is more
complicated because the condition
$[N_\N ,f]_* =0$ allows $f(a,b,\phi,\bar{\phi}|x)$
with $|N_a - N_b| >1 $ that, according to the analysis
of section \ref{$su(2,2)$ - $o(4,2)$ Dictionary}, corresponds to the
higher spin potentials with the symmetry properties of the
$o(4,2)$ Young diagrams having tree rows. Such  fields are
not related to the totally symmetric tensor and tensor-spinor
fields described in \cite{LV,vf} and are expected to correspond to
the mixed symmetry free $AdS_5$ fields which, as shown in
\cite{BMV}, are not equivalent to the symmetric fields in the $AdS_5$
background although becoming equivalent to some their combinations in the
flat limit. Therefore, to proceed towards $AdS_5$ supersymmetric higher spin
gauge theories it is first of all necessary to develop an appropriate
free field formulation of the mixed symmetry fields in the $AdS$
background. This problem is now under study.

Once the formulation of the mixed symmetry fields in $AdS_5$ is
developed, it will allow one to consider higher spin theories
with all $\N$. These theories are expected to be dual to the
$4d$ free supersymmetric conformal higher spin theories analyzed
in \cite{BHS}. In \cite{BHS}  it was  suggested  that
a class of larger $CFT_4$ and $AdS_5$ consistent
higher spin theories should
exist exhibiting manifest $sp(8)$ symmetry. Such theories result
from relaxing the condition that the $AdS_5$ higher spin algebra
is spanned by the elements that commute to $N_\N$.
Being analogous to the $4d$ higher spin gauge theories
based on the algebras $hu(n,m|4)$,
the generalized higher spin gauge theories based on the algebras
$hu(n,m|8)$ are expected to be dual to the $hu(n,m|8)$ invariant
$4d$ conformal higher spin gauge theories \cite{BHS}.
The $5d$ $sp(8)$ invariant
higher spin gauge theories are likely to be generating theories
for the reduced
models based on the centralizers of $N_\N$ in $hu(n,m|8)$
as the ones discussed in this paper. It is tempting to
speculate that the reduction of the higher spin
algebras $hu(n,m|8)$
is a result of a certain spontaneous symmetry breaking mechanism
with some dynamical field $\varphi$ in the adjoint representation of the
higher spin algebra that develops a vacuum expectation value
$\varphi = N_\N +\ldots$.

\vskip0.5cm

{\bf Acknowledgments.}
I am grateful to Kostya Alkalaev  for careful reading the manuscript
and useful comments and to Lars Brink for the hospitality at the
Institute for Theoretical
Physics, Chalmers University, where some essential part of this work
was done.
This research was supported in part by INTAS, Grant No.99-1-590,
the RFBR Grant
No.99-02-16207 and the RFBR Grant No.01-02-30024.

\appendix
\section{Free Field Equations}
\label{Free Field Equations}

In this appendix we give some details on the derivation of
the equations of motion that follow
from the quadratic part of the higher spin action.  The variation
(\ref{deltaS2}) can be equivalently rewritten as
\bee
\label{deltaS21}
\delta S_2 =  -\half \int_{M^5} &{}& \ls \rho (p+q) \Big (
\Big [a_1^\gga \f{\p}{\p b_1^\gga },
\frac{\partial^2}{\partial a_{1\ga}\partial a_{2\gb }}
\bar{c}_{12}\Big ]
+
\Big [a_2^\gga \f{\p}{\p b_2^\gga },
\frac{\partial^2}{\partial a_{1\ga}\partial a_{2\gb }}
\bar{c}_{21}]\Big )
\nn\\
&{}& H_{2\,\ga\gb} \wedge tr ( \delta \go(a_1 ,b_1 |x )
\wedge R_1 (a_2 ,b_2 |x ) ) \Big |_{a_i = b_j =0} \,.
\eee
Taking into account that the Young symmetrizers
$a_1^\gga \f{\p}{\p b_1^\gga }$ and $a_2^\gga \f{\p}{\p b_2^\gga }$
commute to $p+q$ and using the definition of the
component fields (\ref{YD}), (\ref{lgf}) we find
\bee
\label{deltaS22}
\delta S_2 =  -\int_{M^5} &{}& \ls \rho (p+q) tr \Big (
 \frac{\partial^2}{\partial a_{1\ga}\partial a_{2\gb }}
\bar{c}_{12} \delta \eta^1 (a_1 ,b_1 |x )
\wedge R_1 (a_2 ,b_2 |x )\nn\\
\ls\ls&+&
      \frac{\partial^2}{\partial a_{1\ga}\partial a_{2\gb }}
\bar{c}_{21} \delta \go(a_1 ,b_1 |x )
\wedge r^1_1 (a_2 ,b_2 |x ) \Big )
\wedge H_{2\,\ga\gb} |_{a_i = b_j =0} \,.
\eee
Now one observes that the terms containing simultaneously
$\eta^1$ and $r^1_1$
cancel out because $H_{2\,\ga\gb}$ is symmetric
in the spinor indices.
Therefore
\bee
\label{deltaS23}
\delta S_2 =  -\int_{M^5} &{}& \ls \rho (p+q ) \bar{c}_{12}
\frac{\partial^2}{\partial a_{1\ga}\partial a_{2\gb }}
H_{2\,\ga\gb}\nn\\
&{}&\ls\ls\ls\ls\ls\ls \wedge
tr \Big ( \delta \eta^1 (a_1 ,b_1 |x )\wedge r^0_1 (a_2 ,b_2 |x )
+ r^1_1 (a_1 ,b_1 |x )
\wedge \delta \eta^0 (a_2 ,b_2 |x )\Big )
\Big |_{a_i = b_j =0}\,.
\eee
Inserting here the component expansions
\be
\label{ceta}
\eta^i(a,b|x )= \sum_{u,v=0}^\infty \frac{1}{u!v!} \delta (2i - u+v)
\eta^{i\,\ga_1 \ldots \ga_u}{}_{\gb_1\ldots \gb_v} (x) a_{\ga_1}\ldots a_{\ga_u}
b^{\gb_1}\ldots b^{\gb_v}\,,
\ee
\be
\label{cr}
r^i_1 (a,b|x )= \sum_{u,v=0}^\infty \frac{1}{u!v!}\delta (2i - u+v)
r_1^{i\,\ga_1 \ldots \ga_u}{}_{\gb_1\ldots \gb_v}(x)
a_{\ga_1}\ldots a_{\ga_u} b^{\gb_1}\ldots b^{\gb_v}\,,
\ee
\be
\rho (p) = \sum_{n=0}^\infty \f{\rho_n}{n!} p^n\,,
\ee
where, taking into account (\ref{ggac}) and (\ref{g+}),
\be
\label{rn}
\rho_n = (-1)^n \f{1}{2^{2n+2}\,(n+1)!}\,,
\ee
and completing the differentiations one gets
\bee
\label{deltaS24}
\delta S_2 =  -\int_{M^5} &{}& \ls \sum_{u,v=0}^\infty
\f{\rho_{u+v}}{u!v!} (-1)^u H_{2\,\ga\gb}\wedge\nn\\
&{}&\ls\ls\ls tr \Big (
 \delta \eta^{1\ga\gga_1\ldots \gga_{u+1}\kappa_1 \ldots \kappa_{v}}
{}_,{}^{\gs_1\ldots \gs_{u}\rho_1 \ldots \rho_{v}}
\wedge r_1^0{}^\gb{}_{\gs_1\ldots \gs_{u}\kappa_1 \ldots \kappa_{v},
\gga_1\ldots \gga_{u+1}\rho_1 \ldots \rho_{v}}\nn\\
&{}&\ls\ls\ls+
 r_1^{1\ga\gga_1\ldots \gga_{u+1}\kappa_1 \ldots \kappa_{v}}
{}_,{}^{\gs_1\ldots \gs_{u}\rho_1 \ldots \rho_{v}}
\wedge \delta \eta^0{}^\gb{}_{\gs_1\ldots \gs_{u}\kappa_1 \ldots \kappa_{v},
\gga_1\ldots \gga_{u+1}\rho_1 \ldots \rho_{v}}
\Big )
\Big |_{a_i = b_j =0}\,.
\eee
Using (\ref{rn}) and the Young properties of the component fields and
curvatures one obtains
\bee
\label{deltaScom}
\delta S_2 =- \half\phi_0&{}&\ls \int_{M^5}   \sum_{n=0}^\infty
(-1)^n 2^{-2n} \f{1}{(n-1)! n!} H_{2\,\ga\gb}\wedge \nn\\
&\times&tr \Big (
 \delta \eta^{1\ga\gga_1\ldots \gga_{n}}
{}_,{}^{\gs_1\ldots \gs_{n-1}}
\wedge r^0_1{}_{\gga_1\ldots \gga_{n},}{}^\gb{}
{}_{\gs_1\ldots \gs_{n-1}}\nn\\
&{}&+
r^1_1{}^{\ga\gga_1\ldots \gga_{n}}
{}_,{}^{\gs_1\ldots \gs_{n-1}}
\wedge \delta \eta^0{}_{\gga_1\ldots \gga_{n},}{}^\gb{}
{}_{\gs_1\ldots \gs_{n-1}}
\Big )
\Big |_{a_i = b_j =0}\,.
\eee

The free equations of motion corresponding to the variation
with respect to Lorentz-type fields $\eta^1$ and the frame-type
field $\eta^0$ have the following component form, respectively,
\be
\label{htor}
0=H_{\ga_1\gs} r_1^0{}_{\ga_2 \ldots \ga_{m+1},}{}^\gs{}{}_{\gb_1\ldots
\gb_{m-1}} -\f{m-1}{2(m+1)} V_{\ga_1\gb_1} H_{\gga\gs}
r_1^0{}_{\ga_2 \ldots
\ga_{m+1},\gb_2 \ldots \gb_{m-1}}{}^{\gga\gs}
\ee
and
\bee
\label{heq} 0&=& H_{\ga_1}{}^\gga
(r_1^1{}_{\gb_1\ldots\gb_m \gga ,\ga_2 \ldots
\ga_{m}} + \f{m-1}{3}r_1^1{}_{\gb_1\ldots\gb_m \ga_2 ,\ga_3 \ldots
\ga_{m}\gga})\nn\\
&-&mH^\gga{}_{\gb_1}
(r_1^1{}_{\gb_2\ldots\gb_{m} \ga_1\gga
,\ga_2 \ldots \ga_{m}} + \f{m-1}{3}
r_1^1{}_{\gb_2\ldots\gb_m \ga_1\ga_2 ,\ga_3
\ldots \ga_{m}\gga}) \nn\\
&+&\f{m}{m+1} V_{\ga_1\gb_2}H^{\gga\gs}
(r_1^1{}_{\gb_2 \ldots \gb_m \gga\gs,\ga_2\ldots \ga_m}
+(m-1) r_1^1{}_{\gb_2
\ldots \gb_m \gga\ga_2,\gs\ga_3\ldots \ga_m}\nn\\
&+&\,\,\,\,\,\,\,\,\,\,\f{1}{6}(m-1)(m-2)
r_1^1{}_{\gb_2 \ldots \gb_m \ga_2 \ga_3,\gs\gga\ga_4\ldots \ga_m} )\,.
\eee
(As usual, the symmetrization of the indices denoted by the same
Greek letters is assumed).

\end{document}